\documentclass[aps,amsmath,amssymb]{revtex4}
\usepackage[dvips]{graphicx,color}
\usepackage{times}
\usepackage{mathrsfs}
\usepackage{amsmath}
\usepackage{graphicx}
\usepackage{epsfig}
\usepackage{dcolumn}
\usepackage{bm}
\begin{document}
\title{Formation of caustics in Dirac-Born-Infeld type scalar field systems}
\author{ U. D. Goswami$^1$, H. Nandan$^2$ and M. Sami$^2${\footnote{Senior 
associate, Abdus Salam International Centre for Theoretical Physics (ICTP), 
Trieste, Italy.}}}
\affiliation{$1.$ Department of Physics, Dibrugarh University,
Dibrugarh 786004, Assam, India,}
 \affiliation{$2.$ Center for
Theoretical Physics, Jamia Millia Islamia, Jamia Nagar, New
Delhi-110092, India.}
\begin{abstract}
We investigate the formation of caustics in the Dirac-Born-Infeld
type scalar field systems for generic classes of potentials, viz.,
massive rolling scalar with potential, $V(\phi)=V_0e^{\pm \frac{1}{2}
M^2 \phi^2}$ and inverse power-law potentials with
$V(\phi)=V_0/\phi^n,~0<n<2$. It is found that in the case of\texttt{}
exponentially decreasing rolling massive scalar field potential,
there are multi-valued regions and regions of likely to be caustics
in the field configuration. However there are no caustics in the
case of exponentially increasing potential. We show that the
formation of caustics  is inevitable for the inverse power-law
potentials under consideration in  Minkowski space time whereas
caustics do not form in this case in the FRW universe.
\end{abstract}

\maketitle
\section{Introduction}
The discovery of the late time accelerated expansion of universe is
one of the most surprising findings in modern cosmology\cite{Riess}
and posses a serious challenge to fundamental theories of physics.
The resolution of cosmic acceleration riddle similar to the phenomenon of black body radiation
might unveil new secrets of nature.

It is now widely believed that the late time acceleration
of  universe is due to an exotic form of energy with large negative
pressure known as {\it dark energy} which is the dominant fraction of the
energy content of present universe \cite{review1,review2,review3}.
The simplest dark energy model based upon cosmological constant is
faced with fine tuning problem of an acceptable level. As an
alternative to cosmological constant, a variety of scalar field
models is proposed in recent years to provide a viable explanation
for the phenomenon of late time cosmic acceleration
\cite{Zlatev,Armendariz,Kamenshchik,Padmanabhan,Garousi}. Though
the scalar field models have limited predictive power but
nevertheless can be of interest in case they can exhibit some generic
features allowing to alleviate the fine tuning and coincidence problems or
could be motivated from a fundamental theory of high energy physics.
The Dirac-Born-Infeld (DBI) scalar field model is string
inspired and certainly invites attention\cite{sen}. Unlike generic
quintessence models with tracker solutions, there exists no solution
which can mimic scaling matter/radiation regime in case of the
tachyon field
\cite{samicop,samiothers,allthat,Paddy,staro,Bagla,AF,AL,GZ}. These
models necessarily belong to the class of thawing models: At early
times, the expansion dynamics is governed by the background fluid
whereas the tachyon field remains subdominant and frozen. However,
as the background energy density redshifts and becomes comparable to
the field energy density, the field begins to evolve and
subsequently overtakes the background to become the dominant
component of universe.

Tachyon models do admit scaling solution in presence of a
hypothetical barotropic fluid with negative equation of state.
Tachyon fields can be classified by the asymptotic behavior of their
potentials for large values of the field: (i) $V(\phi) \to 0$ faster
than $1/\phi^2$ for $\phi \to \infty$. In this case dark matter like
solution is a late time attractor. Dark energy may arise in this
case as a transient phenomenon. (ii) $V(\phi) \to 0$ slower than
$1/\phi^2$ for $\phi \to \infty$ ; these models give rise to dark
energy as late time attractor. The two classes are separated by $
V(\phi)\sim 1/\phi^2$ which is a scaling potential with
$w({\phi})=constant$. The present state of observations does not allow
to distinguish amongst various scalar field models and leaves the
dark energy metamorphosis  as a future challenge for observational
cosmology.

The formation of caustics in field profile in the mass free space is an
undesirable consequence in the field theoretical models in cosmology
as it indicates the failure of physical theories to explain the
evolution of field in that particular region. Thus the study of
formation of caustics in the field configuration is one of the best techniques 
to investigate the fundamental shortcoming of the field theory for a
specific potential.

Inspite of the exiting features of cosmological dynamics based upon
DBI scalars, it might happen that these models lead to formation
of caustics where the second and the higher-order derivatives of the
field become singular. As demonstrated in
Ref.~\cite{staro}, caustics inevitably form in tachyon system with
potentials decaying faster than $\phi^{-2}$ at infinity. We do not know whether caustics are generic
prediction of string theory or appear as a result of the derivative
truncation leading to the DBI action. It remains
to extend the analysis of Ref.~\cite{staro} to the case of
inverse power law potentials, $V \sim \phi^{-n}$ with $0<n<2$ analysed in
Ref.~\cite{Copeland}. Caustics normally form in systems with
pressureless dust which is mimicked by tachyon field with run away
potentials. It is therefore quite likely that caustics may not
develop in Born-Infeld systems with a ground state at a finite value
of the field. The rolling massive scalar potential, $V(\phi)=V_0
e^{\frac12 M^2\phi^2}$ belongs to this category.

In this paper, we address the issue of caustic formation for DBI
scalar field systems for inverse power potentials giving rise to
dark energy as a late time attractor and the massive rolling scalar.
In the section II, we present the DBI type general scalar field
equation for the 1 + 1 flat space time and review the formalism of
field dynamics related to caustic formation based upon Ref.~\cite{staro}.
We also extend the
analysis to the case of isotopic and homogeneous expanding universe to
see the formation of caustics in the field in more real situation.
In section III, we discuss our numerical results. To have a
comprehensive view of the scalar field pattern, the general field
equations are solved numerically as well analytically in the
homogeneous and inhomogeneous situations for the two
different classes of field potentials mentioned above. For the
inhomogeneous case, we solve the evolution equations analytically using the
method of characteristics \cite{staro}  to check the formation of
caustics in the field profile using the
geometrical interpretation of the solution. Main results of our
analysis are summarized in the last section.
\begin{figure}[hbt]
\centerline
\centerline{\includegraphics[scale=0.25]{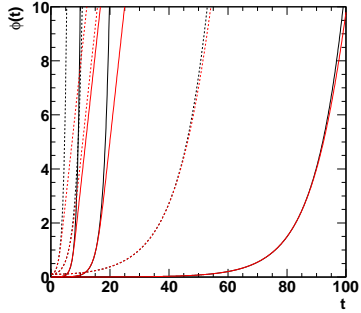}}
\caption{The numerical solution of the equation (\ref{eq19}) (red
line) and the analytic solution of equation (\ref{eq20}) (black line) for
different values of $M$ and $\phi_0$.
The solid line is for the initial value of field $\phi_0$ = 0.001 and the 
dotted line is for  $\phi_0 = 0.1$. Three different sets are due to three 
different values of $M$ = 1, 0.5 and 0.1 along the increasing values of t. 
The discrepancy between these two solutions is obvious after some 
initial period depending upon the values of $M$ and $\phi_0$.} \label{fig1}
\end{figure}
\section{The Dirac-Born-Infeld type scalar field system}
The DBI type action for a  scalar field $\phi$, referred to as
tachyon field hereafter, is given by \cite{sen,staro,review2},
\begin{equation}
S = - \int  d^{4}x \sqrt{-g}~  V(\phi)  \sqrt{1 +
{\partial_{\mu}\phi\partial^{\mu}\phi}}~, \label{eq1}
\end{equation}
where $V(\phi)$ is the potential of the field $\phi$. The field
equation derived from the action (\ref{eq1}) is,
\begin{equation}
\nabla_{\mu}\nabla^{\mu}\phi - \frac{\nabla_{\mu}\nabla_{\mu}\phi}{1
+ \nabla_{\alpha}\phi\nabla^{\alpha}\phi}
\nabla^{\mu}\phi\nabla^{\nu}\phi - \frac{V_{,\phi}(\phi)}{V(\phi)} = 0,
\label{eq2}
\end{equation}
where the covariant derivative of the field $\phi$ with respect to
the metric $g_{\mu\nu}$ is denoted by $\nabla_{\mu}$.
\subsection{Minkowski 1+1 dimensional analysis}
In what follows, we shall first consider the field equations in 1 +
1 dimensional Minkowski space. Caustics formation can easily be
analysed in this case. If caustics do not form in Minkowski space
time, we would expect the result to hold in the expanding universe
as expansion should work against the formation of caustics. However,
in the opposite case, it essential to incorporate expansion to reach
the final conclusion.

In this case, the equation (\ref{eq2}) takes the form,
\begin{equation}
\ddot{\phi} = \left(1 + \phi^{\prime 2}\right)^{-1}\left[\left(1-
\dot{\phi}^2 + \phi^{\prime 2}\right)\left(\phi^{\prime \prime} -
\frac{V_{,\phi}(\phi)}{V(\phi)}\right) +
2\phi^{\prime}\dot{\phi}\dot{\phi}^{\prime} - \phi^{\prime
2}\phi^{\prime \prime}\right]. \label{eq3}
\end{equation}
Here the time and space derivatives of the field $\phi$ are
indicated by the dot and prime over $\phi$ respectively.

In order to understand the time and space evolution of the  field
$\phi$ in 1 + 1 dimension under a specific field potential, let us
first consider some general aspects related to the equation
(\ref{eq3}) for the homogeneous and inhomogeneous fields.

If the scalar field is homogeneous, then $\phi^{\prime} = 0$, and
hence the equation (\ref{eq3}) for such a field simplifies
to,
\begin{equation}
\frac{\ddot{\phi}}{1 - \dot{\phi}^2} + \frac{V_{,\phi}(\phi)}{V(\phi)}
= 0. \label{eq4}
\end{equation}
For a particular potential, the time evolution of the homogeneous
field can be obtained from this equation (\ref{eq4}). In the case of
an inhomogeneous scalar field, $\phi^{\prime}$, $\phi^{\prime\prime}
> 0$, however they should be sufficiently less then unity for a
realistic scalar field in cosmology \cite{staro}. Under this
condition, the equation (\ref{eq3}) can be written as,
\begin{equation}
\ddot{\phi} = \left(1- \dot{\phi}^2 + \phi^{\prime
2}\right)\left(\phi^{\prime \prime} -
\frac{V_{,\phi}(\phi)}{V(\phi)}\right) = P(\phi)Q(\phi), \label{eq5}
\end{equation}
where,
\begin{equation} \label{eq6}
\begin{split}
P(\phi) = 1- \dot{\phi}^2 + \phi^{\prime 2}.\\
Q(\phi) = \phi^{\prime \prime} - \frac{V_{,\phi}(\phi)}{V(\phi)}.
\end{split}
\end{equation}
The equation of $P(\phi)$ contains both the time and space
derivatives of the field and therefore this parameter can be used as
an indicator of the pattern of evolution of the field governed by
the particular field potential. For example, if the field $\phi(x,
t)$ rapidly approaches  the configuration $P(\phi) = 0$,  it then
indicates that the time and space evolution of the field is such
that it rapidly approaches  $\dot{\phi}^2 - \phi^{\prime 2} = 1$.
Alternatively, we may say that, under this situation the field
$\phi(x, t)$ is almost similar to some subsidiary field $\theta(x
,t)$ whose time and space evolution is constrained by the equation
\cite{staro},
\begin{equation} \label{eq7}
\dot{\theta}^2 - \theta^{\prime 2} = 1.
\end{equation}
The field may approaches to the configuration $P(\phi) = 0$ from both 
directions, viz. from above or below zero, depending upon the initial state
of the field. If the initial state of the field is such that initially 
$P(\phi) > 0$, then $P(\phi)$ will asymptotically approach to zero from above,
otherwise it will tend to be zero from below \cite{staro}.    
Similarly, if $P(\phi) \approx 1$, it implies that, $\dot{\phi}^2
\approx \phi^{\prime 2}$, i.e., the type of evolution of the field
$\phi(x, t)$ with respect to space and time are nearly equal. In
this case also, we may consider that the $\phi(x, t)$ is almost
similar to some subsidiary field $\omega(x ,t)$ whose time and space
variations are related by the equation,
\begin{equation} \label{eq8}
\dot{\omega}^2 - \omega^{\prime 2} = 0.
\end{equation}
The extensive numerical solutions of the equation (\ref{eq5}) for
the scalar field with the two different classes of field potentials
of our interest clearly showed the above  behavior of $P(\phi)$.
Thus the equation (\ref{eq5}) has two attractors, one for $P(\phi) =
0$ and the other for $P(\phi) = 1$, depending on the field potential
and hence $P(\phi)$ can be used to define the following relation,
\begin{equation}
\dot{\phi} = \sqrt{1 + \phi^{\prime 2} - P(\phi)}. \label{eq9}
\end{equation}
\begin{figure}[hbt]
\centerline
\centerline{\includegraphics[scale=0.25]{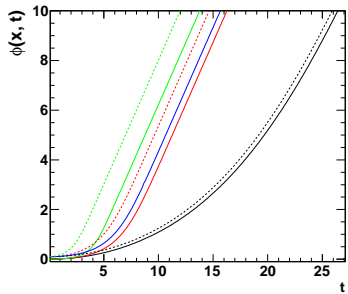}}
\caption{The numerical solutions of the equation (\ref{eq22}) for
different values of $M$ and $\phi_0$.  Black, red and
green lines are for $M$ = 0.1, 0.5 and 1 respectively. The solid line 
represents $\phi_0$ = 0 and the dotted line $\phi_0$ = 0.1. For all these
plots we consider $\phi^{\prime}$ = 0.01 and $\phi^{\prime\prime}$ = 0.02. 
The blue line represents the solution with $\phi_0$ = 0, $M$ = 0.5, 
$\phi^{\prime}$ = 0.002 and $\phi^{\prime\prime}$ = 0.001. For all cases 
time variation of the field is very slow initially and then it increases 
rapidly as time passages.} \label{fig2}
\end{figure}
\begin{figure}[hbt]
\centerline
\centerline{\includegraphics[scale=0.25]{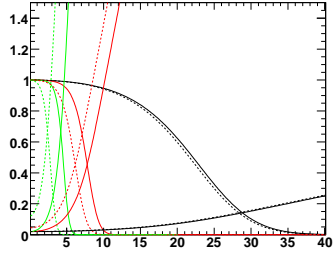}}
\caption{The variation patterns of $P(\phi)$ and $Q(\phi)$ with
respect to time for the potential (\ref{eq18}) corresponding to
the initial conditions of the figure \ref{fig2}. The falling lines are for 
$P(\phi)$ and raising lines are for $Q(\phi)$. The overall pattern are same 
for all initial conditions.} \label{fig3}
\end{figure}
If we consider that the scalar field $\phi(x, t)$ itself satisfies
the subsidiary field equations (\ref{eq7}) and (\ref{eq8}) under
different conditions, then analytic solutions of these two equations
correspond to free relativistic massive wave propagation along the
characteristics of the field $\phi(x, t)$. From the particle point
of view, equations (\ref{eq7}) and (\ref{eq8}) can be viewed as the
dynamical descriptions of the motion of free massive particle under
different situations. From this point of view, the trajectories
$x(t, q)$, (where $q$ is the initial spatial coordinate of the
field, i. e. $x(0,q) = q$) of individual particles and the
corresponding evolution of the field configuration can be obtain
from the parameterized solutions of the characteristic equations of
two variables, viz., $q$ and the affine parameter $s$ along each
curve, which are given by \cite{staro},
\begin{equation}
\frac{dx}{ds}(q, s) = P,\phi^{\prime}  = 2\phi^{\prime};\;\; x(q,0)
= q. \label{eq10}
\end{equation}
\begin{equation}
\frac{dt}{ds}(q,s) = P,\dot{\phi} = -2\dot{\phi};\;\; t(q,0) = 0.
\label{eq11}
\end{equation}
\begin{equation}
\frac{d\phi}{ds}(q,s) =\phi^{\prime}P,\phi^{\prime} +
\dot{\phi}P,\dot{\phi} = 2\phi^{\prime 2} - 2\dot{\phi^{2}};\;\;
\phi(q,0) = \phi_{i}(q). \label{eq12}
\end{equation}
Solving these equations and using equation (\ref{eq9}) we may write,
\begin{equation}
x(q, s) = q + 2s\phi_{i,q}. \label{eq13}
\end{equation}
\begin{equation}
t(q,s) = -2s\sqrt{1 + \phi_{i,q}^{2} - P(\phi)}. \label{eq14}
\end{equation}
\begin{equation}
\phi(q,s) = \phi_{i}(q),\; \mbox{for}\; P(\phi) = 1\;\;\mbox{and}
\;\phi(q,s) = \phi_{i}(q) -2s,\; \mbox{for}\; P(\phi) = 0.
\label{eq15}
\end{equation}
Where $\phi_{i,q} = \phi^{\prime}(q,s)$. Eliminating the affine
parameter $s$ from the above equations, we obtain the parametric
solution for the field $\phi(x,t)$ as,
\begin{equation}
x(q, t) = q - \frac{\phi_{i,q}}{\sqrt{1 + \phi_{i,q}^{2} -
P(\phi)}}t. \label{eq16}
\end{equation}
\begin{equation}
\phi(q,t) = \phi_{i}(q),\; \mbox{for}\; P(\phi) = 1\;
\mbox{and}\;\phi(q,t) = \phi_{i}(q) + \frac{t}{\sqrt{1 +
\phi_{i,q}^{2}}},\; \mbox{for}\; P(\phi) = 0. \label{eq17}
\end{equation}
The equation (\ref{eq16}) will provide us the trajectories of the
free massive particles in the field and the equation (\ref{eq17})
will give us the pattern of evolution of the field with respect to
time \cite{staro}. Hence these two equations will give us the
geometrical form to see the formation of caustics in the field profile and
the evolution of the field configuration with any presumed initial
field profile under a specific field potential. In what follows, we
shall consider the potentials of our interest for the specific
solutions of the equations (\ref{eq4}) and (\ref{eq5}). Although our
main interest is the solution of equation (\ref{eq5}), for the
completeness of the picture we present the solution of equation
(\ref{eq4}) also. From the solutions of equation (\ref{eq5}) for the
following three different field potentials, we shall obtain the
characteristics curves from equation (\ref{eq16}) which will show
clearly whether there are caustics in the field configuration.
\subsection{Generalization to the case of expanding universe}
So far, for simplicity we have considered the field in 1 + 1
dimensional Minkowski space time. However for the real cosmological
applications,  it is necessary to consider the field in 3 + 1
dimensions in the expanding universe. It should be noted that the
extension of the field from 1 + 1 dimension to the 3 + 1 dimensions
is a matter of analogical enhancement of the one dimensional spatial
coordinate to the three dimensional spatial coordinates. Hence the
basic form of all equations discussed above will remain same in 3 +
1 dimensions, also except that the space derivative has to be replaced
by the nabla operator $\nabla$, and the spatial coordinates $x$ and
$q$ by the vectors ${\bf{r}}$ and ${\bf{q}}$ respectively in the
relevant equations. Similarly if we take a spatially flat
Friedmann-Robertson-Walker (FRW) metric with a scale
factor $a(t)$, then equation for the scalar field $\phi$ in an
expanding universe can be obtain from the equation (\ref{eq2}) as,
\begin{equation}
\frac{\ddot{\phi}}{1 - \dot{\phi}^2} + 3H\dot{\phi} +
\frac{V_{,\phi}(\phi)}{V(\phi)} = 0, \label{eq4a}
\end{equation}
where $H = \dot{a}/a$ is the Hubble rate and can be expressed in the
scalar field dominated expanding universe as,
\begin{equation}
H^2 = \frac{8\pi G}{3}\frac{V(\phi)}{\sqrt{1 - \dot{\phi}^2}}.
\label{eq4b}
\end{equation}
The equation (\ref{eq4a}) is equivalent to the equation (\ref{eq4})
of 1 + 1 dimension. If some small inhomogeneous perturbations arise
in the scalar field of the expanding universe, then these
perturbations will generate the small perturbations in the FRW
metric which can be expressed by the Newtonian gravitational
potential $\Phi_{G}(t, {\bf r})$ as \cite{staro,staro1},
\begin{equation}
ds^2 = - (1 + 2\Phi_{G})dt^2 + (1 - 2\Phi_{G})a^2(t)dr^2,
\label{eq5a}
\end{equation}
where we have use the metric convention as $(-,+,+,+)$. If for a
particular potential of the scalar field of expanding universe with
small inhomogeneous field perturbations lead to the state of the
field as in the cases of the equations (\ref{eq7}) and (\ref{eq8})
of the 1 + 1 dimension, then the corresponding equations for the
field $\phi$ itself of the expanding universe represented by the
metric equation (\ref{eq5a}) are given by \cite{staro},
\begin{equation}
\dot{\phi}^2 - \frac{1}{a^2}(\nabla \phi)^2 = 1 + 2\Phi_{G}.
\label{eq5b}
\end{equation}
\begin{equation}
\dot{\phi}^2 - \frac{1}{a^2}(\nabla \phi)^2 = 0. \label{eq5c}
\end{equation}
If we solve these two equations by the method of characteristics as
in the case of 1 + 1 dimension the corresponding equations of
(\ref{eq16}) and (\ref{eq17}) would be,
\begin{equation}
{\bf r}({\bf q}, t) = {\bf q} - \frac{1}{a^2}\frac{(\nabla_{\bf
q}\phi) t}{ \sqrt{1 + \frac{1}{a^2}(\nabla_{\bf q}\phi)^{2} +
2\Phi_{G}}},\; \mbox{for}\; P(\phi) = 0\; \mbox{and}\; {\bf r}({\bf
q}, t) = {\bf q} - \frac{t}{a},\; \mbox{for}\; P(\phi) = 1.
\label{eq5d}
\end{equation}
\begin{equation}
\phi({\bf q},t) = \phi_{i}({\bf q}) + \frac{(1 + 2\Phi_G)t}{\sqrt{1
+ \frac{1}{a^2}(\nabla_{\bf q}\phi)^{2} + 2\Phi_{G}}},\; \mbox{for}\;
P(\phi) = 0\; \mbox{and}\; \phi({\bf q},t) = \phi_i({\bf q}),\;
\mbox{for}\; P(\phi) = 1. \label{eq5e}
\end{equation}
The scale factor $a(t)$ assumes different forms at different stages
of the evolution of universe. For example, in radiation dominated
stage, $a(t) = a_0t^{1/2}$ and in the scalar field dominated stage
$a(t) = a_0t^{2/3}$ \cite{staro,staro1} if the tachyon potential
vanishes at infinity faster than $1/\phi^2$. In case of rolling
massive scalar or the inverse power-law potentials $V(\phi) \sim 1/\phi^n$
with $0<n<2$ , dark energy is a late time attractor and the scale
factor   takes the form $a(t) = a_0t^{2/(3(1+w))}$ with suitable negative
values of $w$. The fluctuations in the tachyon field grow lineally
with time while the metric fluctuations remain constant for
potentials decreasing faster than $1/\phi^2$ at
infinity\cite{staro,staro1}. However, in case of rolling massive
scalar and inverse power-law potentials with $0<n<2$ that would be of interest 
to us, both the fluctuations do not grow \cite{staro1,Garousi}
\section{Caustic formation}
In this section we apply the above formalism to investigate the
possibility of caustic formation in the DBI tachyon system with two
generic classes of potentials and present the results of numerical
simulation.
\subsection{Exponentially decreasing rolling massive potential}
The exponentially decreasing rolling massive scalar field potential
is given by\cite{Copeland},
\begin{equation}
V(\phi) = V_{0}e^{-\frac{1}{2}M^{2}\phi^{2}},
\label{eq18}
\end{equation}
where $V_{0}$ and $M$ are constants.  For this potential the
homogeneous scalar field equation (\ref{eq4}) becomes,
\begin{equation}
\frac{\ddot{\phi}}{1 - \dot{\phi}^2}  - M^{2}\phi = 0.
\label{eq19}
\end{equation}
The numerical solution of this equation is shown in the figure
\ref{fig1} for different values of the arbitrary constant $M$ and initial 
values of field $\phi_0$. The values of $M$ chosen in this figure are 0.1, 
0.5 and 1. On the other hand two values of $\phi_0$ are considered for this 
figure, which are 0.001 and 0.1. If the time variation of the scalar field is 
very slow, then we may consider that $\dot{\phi} << 1$. In this case, equation
(\ref{eq19}) further simplifies to,
\begin{equation}
\ddot{\phi}  - M^{2}\phi = 0.
\label{eq20}
\end{equation}
The solution of this equation is trivial and can be written as,
\begin{equation}
\phi = \phi_{0}cosh(Mt),
\label{eq21}
\end{equation}
where $\phi_{0}$ is the constant initial field value. The plots of this 
equation are also shown in the figure \ref{fig1} for same values of $M$ and 
$\phi_0$ as for the above case to compare with the numerical solution. The red 
line indicates the numerical solution and black line indicates the analytic 
one. The solid line is for the initial field $\phi_0$ = 0.001 and  the dotted 
line is for $\phi_0$  = 0.1. Three different sets are due to three different 
values of $M$ = 1, 0.5 and 0.1 along the increasing values of t.  It is 
observed from the figure \ref{fig1} that the assumption of very slow variation 
of the field with respect to time is correct only for some initial period that 
depends upon the values of $M$ and $\phi_0$. For low values $M$ and $\phi_0$ 
this period is longer than their higher values. As the time 
increases  beyond this period, the discrepancy between the solutions of 
equations (\ref{eq19}) and (\ref{eq20}) increases rapidly. The overall 
trend of time variation of the field with different possible initial conditions 
is almost same, the only difference is that, this 
variation is slower for some initial period if the field is started with 
smaller values of $M$ and $\phi_0$ than their higher values. 

The inhomogeneous 
scalar field equation(\ref{eq5}) for this potential can be written as,
\begin{equation}
\ddot{\phi} = \left(1 - \dot{\phi}^{2} + \phi^{\prime 2}\right)\left(\phi^{\prime\prime} + M^{2}\phi\right).
\label{eq22}
\end{equation}
The numerical solutions of the equation (\ref{eq22}) for the same values of
$M$ as in the previous cases, but for $\phi_0$ = 0 and 0.1 are shown in the 
figure \ref{fig2}. In the figure the black line is for $M$ = 0.1, red line is 
for $M$ = 0.5 and green line for $M$ = 1. The solid line represents for 
$\phi_0$ = 0 and the dotted line for $\phi_0$ = 0.1. Since the solution of the
inhomogeneous equation (\ref{eq22}) depends upon the values of $\phi^{\prime}$
and $\phi^{\prime\prime}$, so for these plots in this figure we consider
$\phi^{\prime}$ = 0.01 and $\phi^{\prime\prime}$ = 0.02. Moreover to check
the dependency of the solution on $\phi^{\prime}$ and $\phi^{\prime\prime}$, we
consider another set of values of these two parameters as 0.002 and 0.001 
respectively together with $M$ = 0.5 and $\phi_0$ = 0.1. The result of this
solution is shown in this figure by the blue line. We observe that for all 
cases the variation or the evolution of the field with time is very slow in 
the initial period based on the values of the $M$ and $\phi_0$, then it 
develops rapidly as time passages. For low values of $M$ and $\phi_0$ this
initial period enhances slightly and the time variation is
also relatively slower for such values. The value of $M$ has significant effect
over the variation of the field with time than $\phi_0$ which is also noticed
for the homogeneous case. Since this is an arbitrary parameter, we can assume
any value of it for a solution, however its actual value can not be 
substantially small because for such a small value the time evolution of the
field will halt for a considerable period of time. The $\phi^{\prime}$ and 
$\phi^{\prime\prime}$ does not have any effect on the time evolution of the 
field, whereas they only effect to shift the magnitude of field in the same 
direction of their magnitude variations (i.e. along the increasing or 
decreasing order). Thus for our remaining discussion, related to this field
potential, we will not consider again 
the effect $\phi^{\prime}$ and $\phi^{\prime\prime}$ over the field.     

The variations of $P(\phi)$ and $Q(\phi)$ with respect to time that are
obtained from the numerical solutions for $\phi(x ,t)$ are plotted in
the figure \ref{fig3}. In this figure, the falling lines are for $P(\phi)$ and 
raising are
for $Q(\phi)$. The color and style of the line represent the corresponding 
values of $M$ and $\phi_0$ as in the case of the figure \ref{fig2}. As 
mentioned earlier the state of the field with respect to time and space is 
indicated by the behavior of the $P(\phi)$ at the corresponding time. so, here 
we are mainly interested in $P(\phi)$ only.

\begin{figure}[hbt]
\centerline
\centerline{\includegraphics[width = 8cm, height = 8cm]{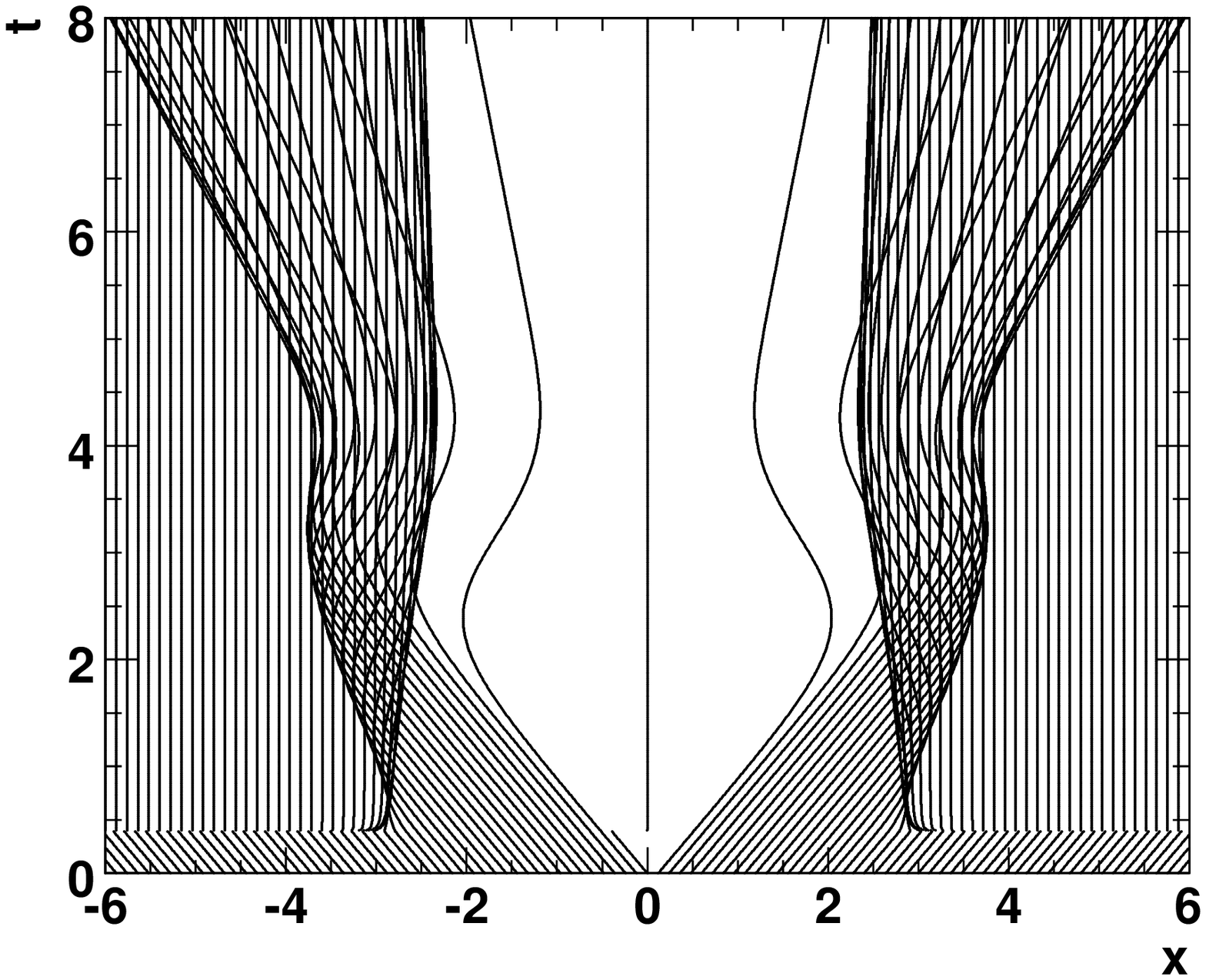}
\includegraphics[width = 8cm, height = 8cm]{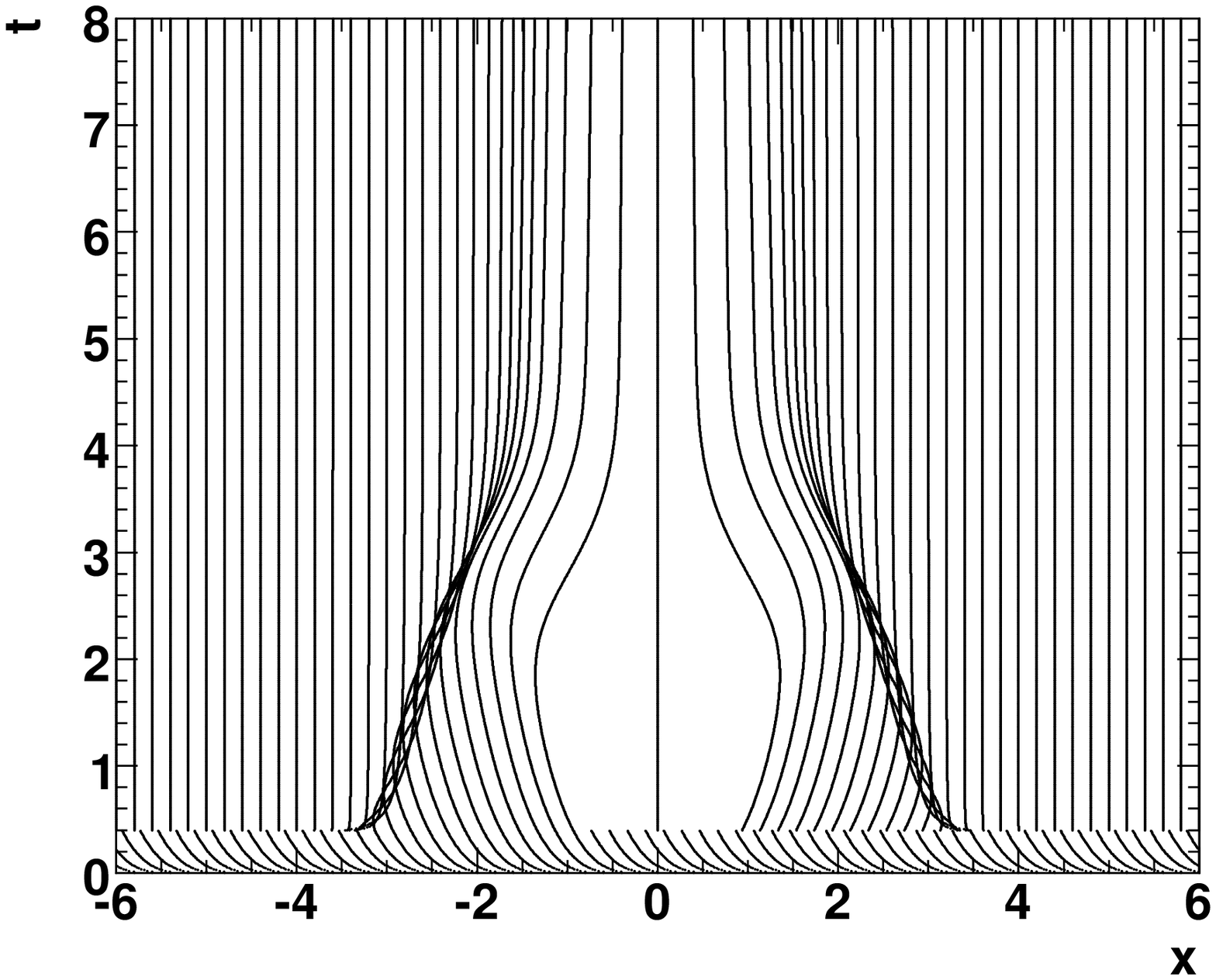}
}
\caption{The particle trajectories in 1 + 1 Minkowski space obtain from 
equation (\ref{eq16}) (left) and in FRW expanding universe obtain from 
equation (\ref{eq5d}) (right) with potential (\ref{eq18}) and 
initial field profile $\phi_{i}(q) = exp(-q^{2})$. Both panels of figure are 
obtain with the initial condition $M$ = 1 and $\phi_0$ = 0.}
\label{fig4}
\end{figure}

From the pattern of variation of $P(\phi)$ with time we observed
that during some initial period, which depends upon the value of $M$ and 
$\phi_0$, the value of $P(\phi)$ is $\approx 1$ and beyond this period its 
value falls rapidly to zero. The length of this initial period is more for 
lower values of $M$ and $\phi_0$ (effect is prominent for the parameter $M$ as
mentioned earlier), otherwise the behaviour of time variation of $P(\phi)$ is 
similar for all initial conditions. To be  more specific, we consider 
the solution of $P(\phi)$ with $M$ = 1 and $\phi_0$ = 0, and the initial field 
configuration as $\phi_{i}(q) = exp(-q^{2})$ \cite{staro}, then we obtain the 
characteristic curves from equation (\ref{eq16}) which are shown in the 
left panel of the figure \ref{fig4}.
From these characteristic curves it is interesting to note that
there are likely to be caustics as well as multi-valued regions in the
field profile. Apart from these, there are regions of twisting of the
characteristic curves. The regions of caustics and multi-valued
start at the points $x = \pm 3$. As the behaviour of the $P(\phi)$ is
similar for all initial conditions mentioned above, therefore we will get the
similar characteristic curves with caustics and multi-valued regions at 
different locations corresponding to the values of $M$ and $\phi_0$. Obviously 
for smaller values of $M$ and $\phi_0$ these location will shift to the higher
value of t. But it is sure that there is no way to avoid these unphysical 
regions for any value of initial field parameters. Thus we may infer that
the formation of caustics and multi-valued regions are independent of the 
initial conditions of the field. 

\begin{figure}[hbt]
\centerline
\centerline{\includegraphics[scale=0.25]{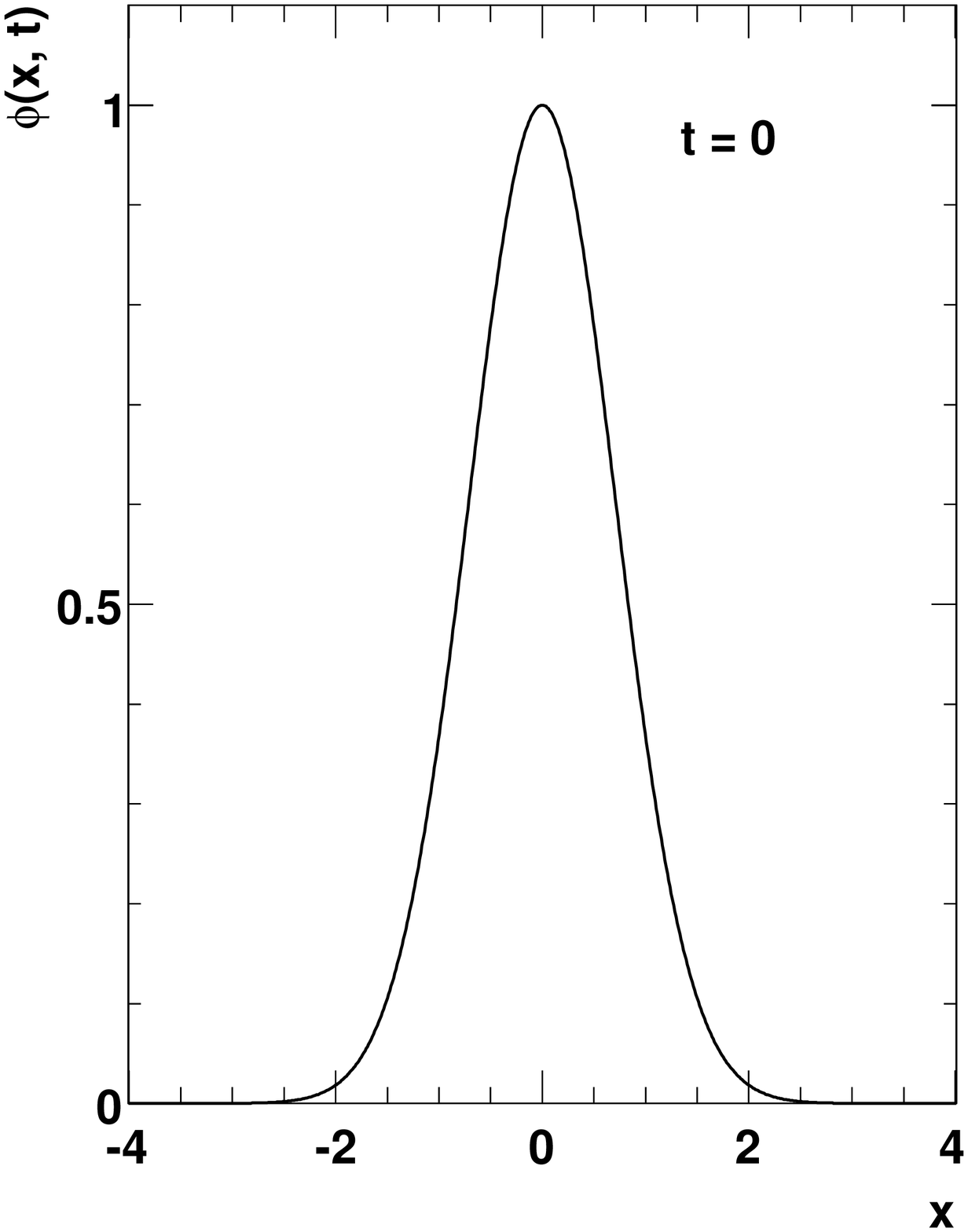}
\includegraphics[scale=0.25]{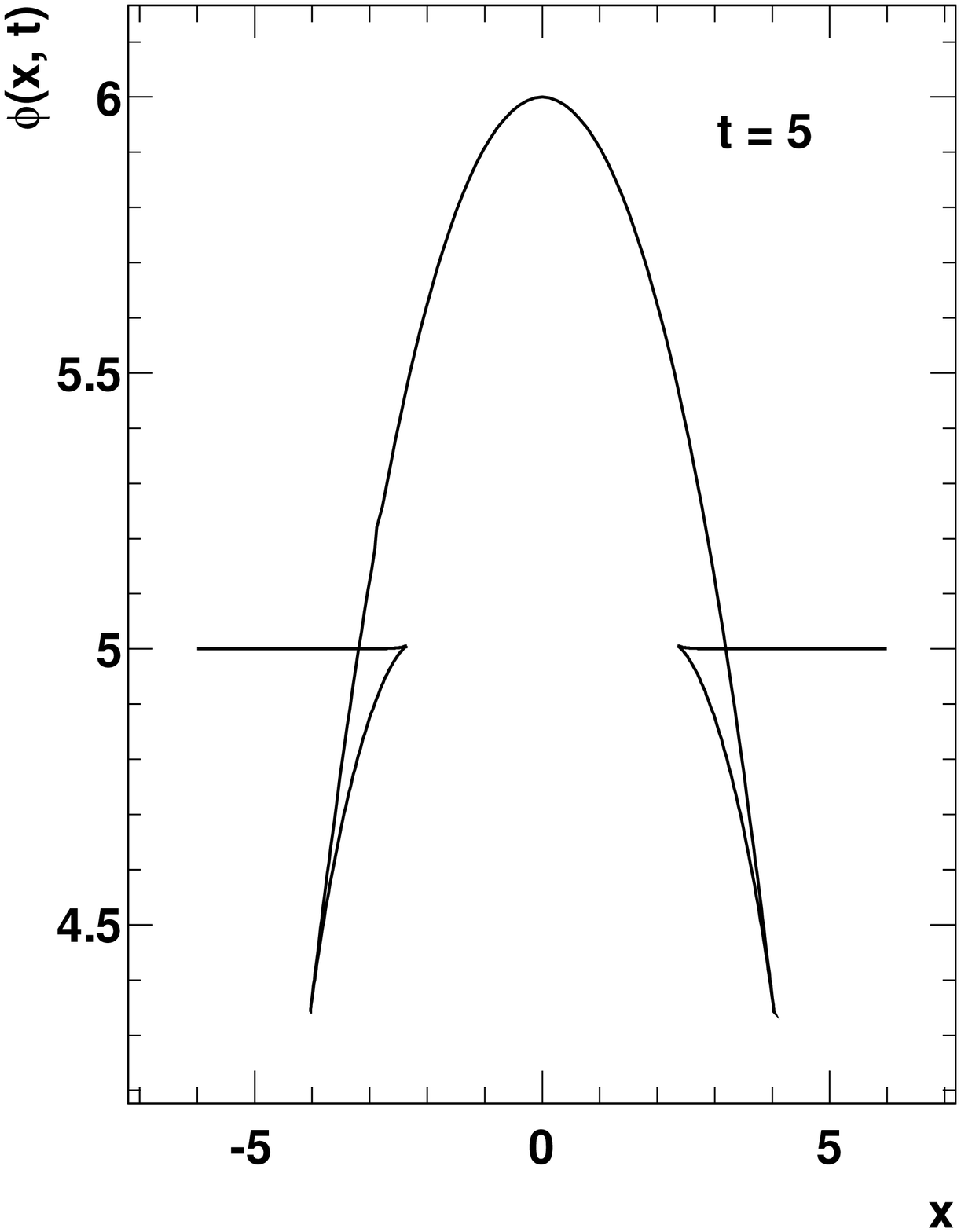}
\includegraphics[scale=0.25]{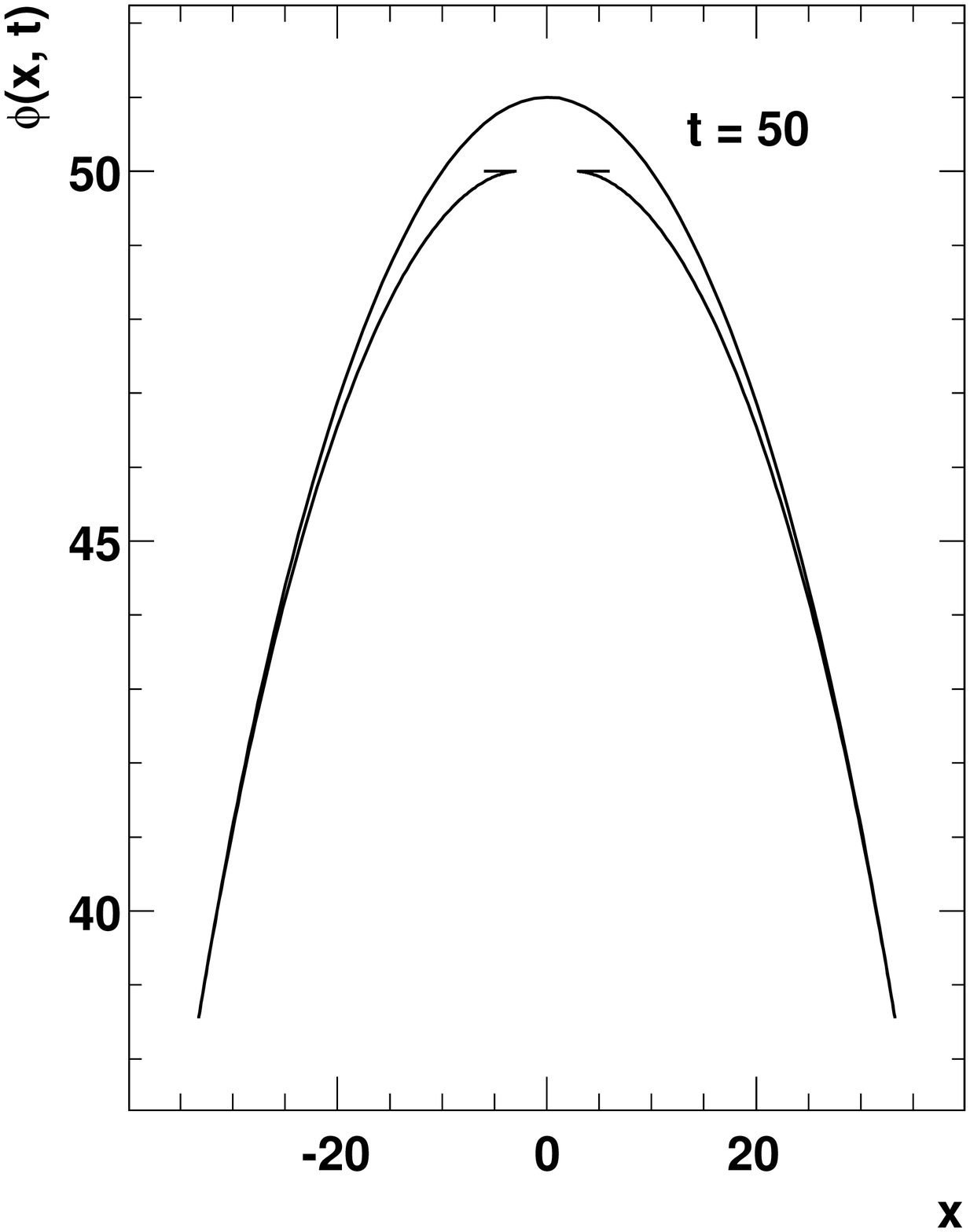}}
\centerline{\includegraphics[scale=0.25]{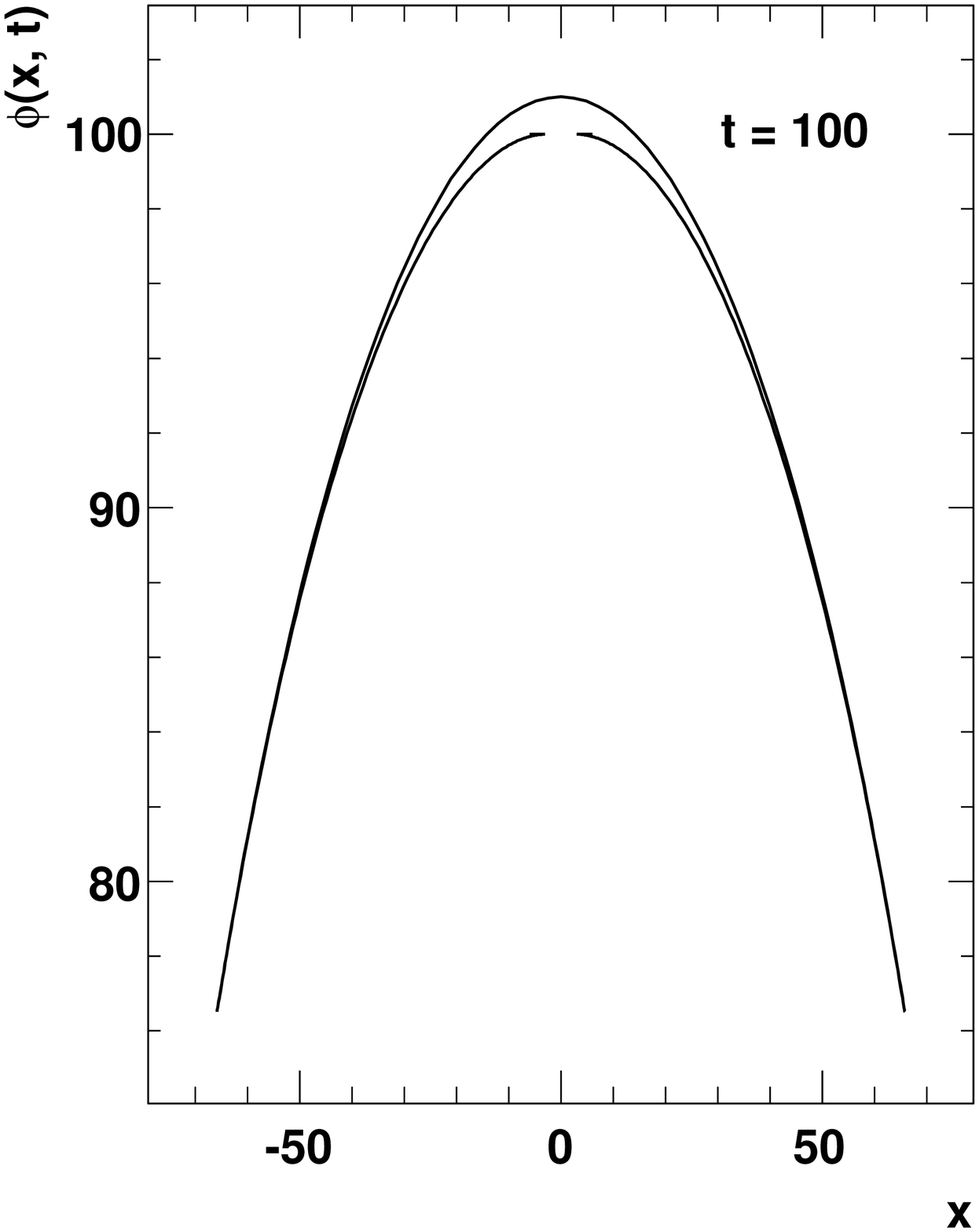}
\includegraphics[scale=0.25]{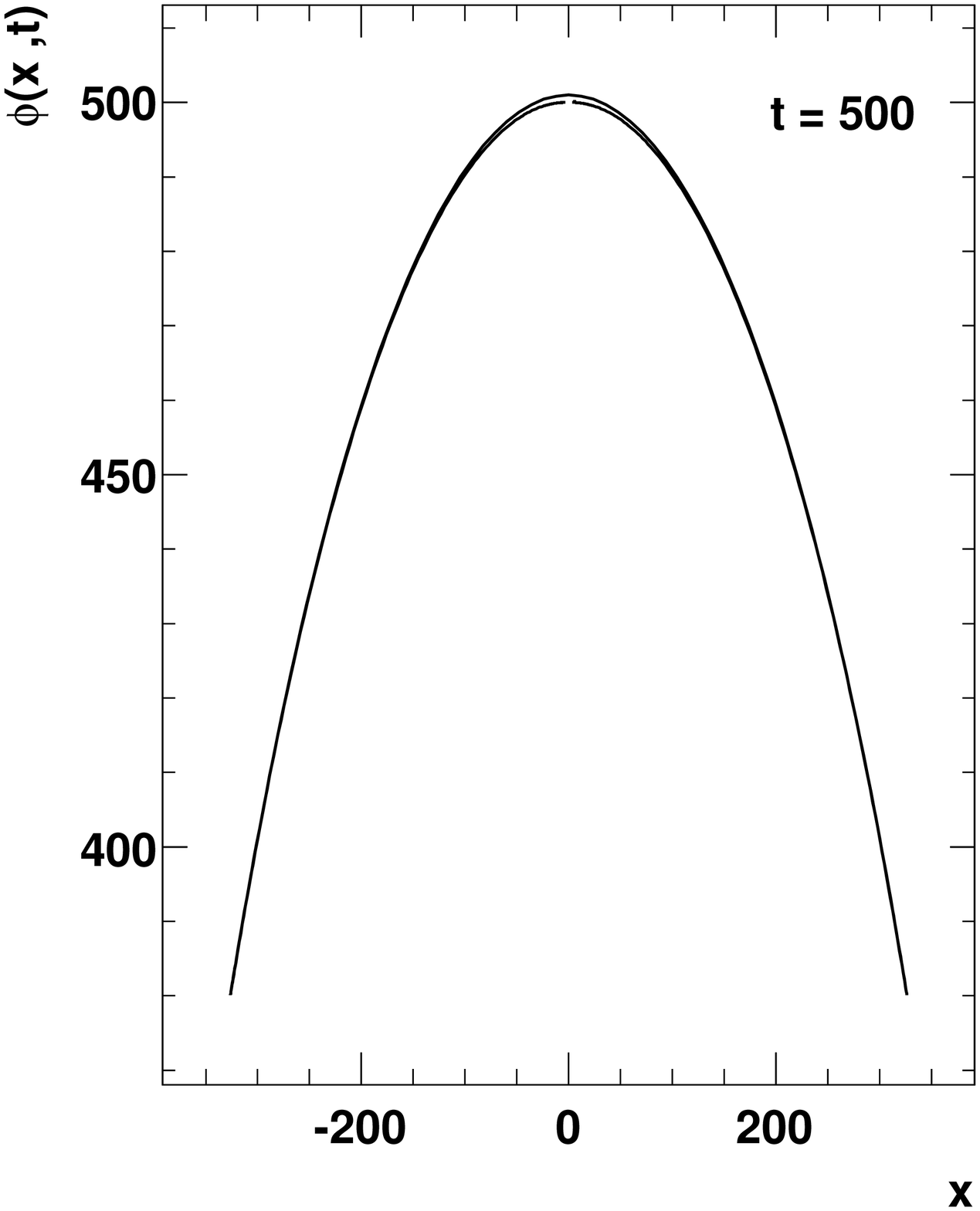}
\includegraphics[scale=0.25]{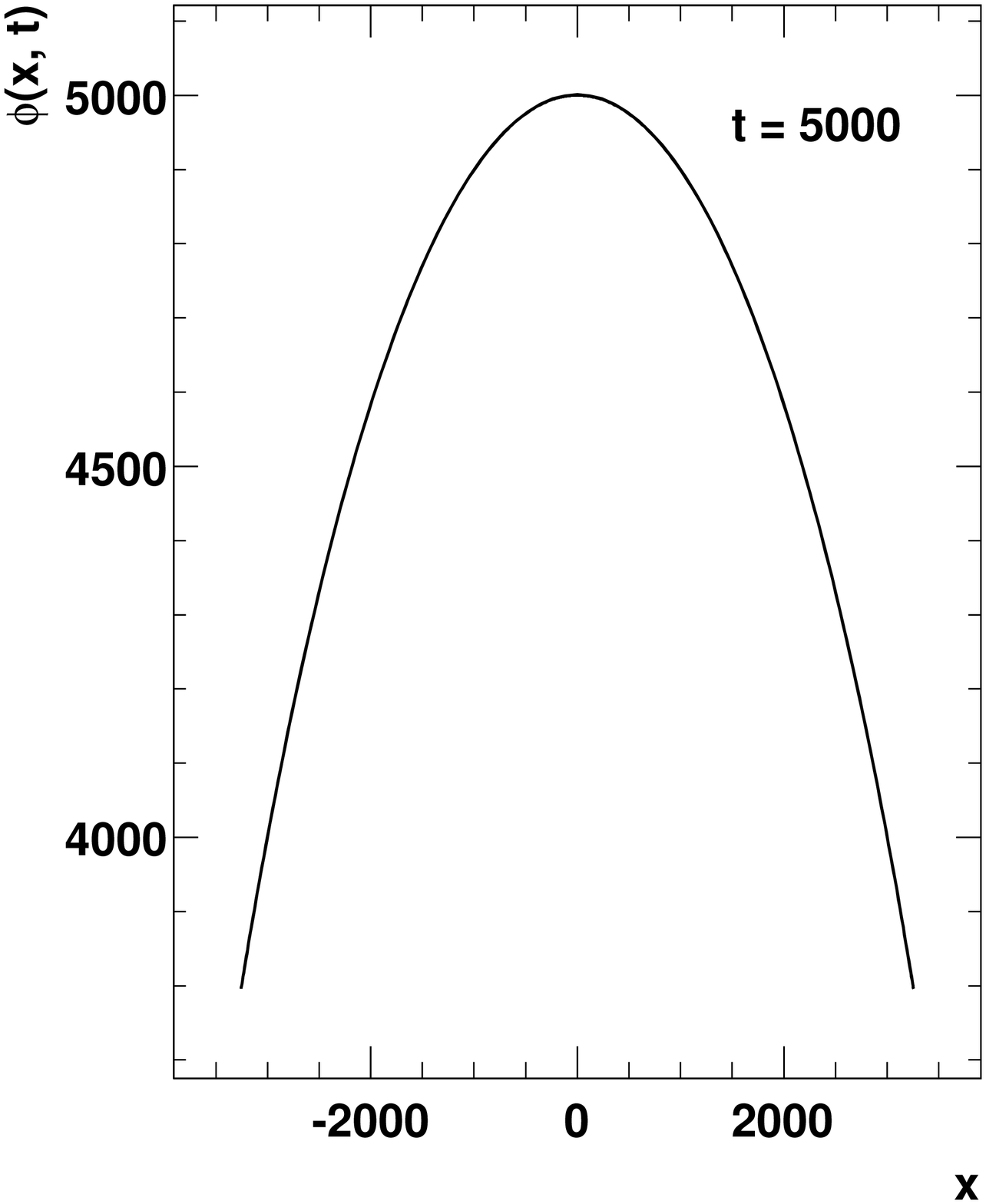}}
\caption{The evolution of the field configuration $\phi(x, t)$ with potential
(\ref{eq18}) and $\phi_{i}(q) = exp(-q^{2})$ at different times for
the initial condition $M$ = 1 and $\phi_0$ = 0.}
\label{fig5}
\end{figure}

The evolution of the field configuration $\phi(x, t)$ with 
$\phi_{i}(q) = exp(-q^{2})$ and for the same initial condition as in the case 
of characteristic curves are shown in the figure \ref{fig5} at different times. 
It should be pointed out that to draw the field configuration at $t = 5$ we 
consider that $P(\phi) = 0$ (which is almost the case. Beyond this value of
$t$, $P(\phi)$ is obviously zero). Within the period, when $P(\phi)$
falls from almost one to zero, field evolves very rapidly with time,
which is not significant to be visualized geometrically. The same observation
can be made for the evolution of the field configuration with different
initial conditions and consequently with the different values of time after 
which $P(\phi)$ falls to zero. 

Heuristically speaking,
expansion works against caustics formation. It is therefore necessary
to investigate the corresponding situation in case of the expanding
universe using equations (\ref{eq5d}) and
(\ref{eq5e}). Without going into detail of the scalar field dynamics in
the expanding universe under this potential, we assume that the
field behaves in a manner equivalent to 1 + 1 dimensional case (which
is specified by the behavior of $P(\phi)$). Since the average matter
density of the universe is quite low to obtain a substantial value
of the Newtonian gravitational potential $\Phi_G(t, {\bf r})$
\cite{Tren,Lee} of the universe, we shall neglect $\Phi_G$ in comparison to 
unity. Secondly, as dust like solution is a late attractor in the present 
case, we assume $a(t) \sim t^{2/3}$. Under these considerations,  we obtain the
characteristic curves from equation (\ref{eq5d}) with the initial
field profile as mentioned above are shown in the right panel of the
figure \ref{fig4} for the one dimensional space. It is observed that
the patterns of the characteristics curves are different in this
case from the 1 + 1 dimensional Minkowski space time and the
caustics are more distinctly formed in the field profile if the
field behave in a similar manner as in the 1 + 1 dimension with this
exponentially decreasing rolling massive potential. From the
numerical calculations under the above considerations we found that
the evolution patterns of the field obtained from the equation
(\ref{eq5e}) remain nearly same as its initial profile. We should note that
the field rolls slowly for a short while near the origin but quickly
enters the fast roll to mimic dark matter like regime described by 
$a(t) \sim t^{2/3}$. Using equations (18) and (19), we have verified that 
situation depicted in figure \ref{fig4} does not change if the detailed field 
dynamics is taken into account.

\subsection{Exponentially increasing rolling massive potential}
\begin{figure}[hbt]
\centerline
\centerline{\includegraphics[scale=0.5]{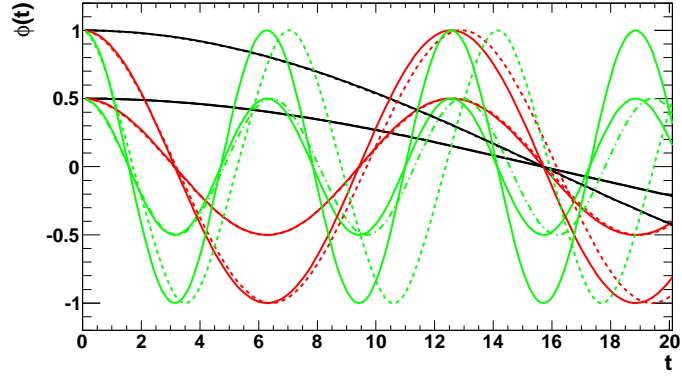}}
\caption{The numerical solutions of the equation (\ref{eq24}) (dotted line) and
the solutions of equation (\ref{eq25}) (solid line) for three different values
of $M$ = 0.1, 0.5 and 1, which are represented by black, red and green colors 
respectively. The upper set is for $\phi_0$ = 1 and lower set is for $\phi_0$ 
= 0.5.  There is a good agreement between the numerical and analytic solutions 
although there is gradual phase shift between them as time passages depending
on the values of $M$ and $\phi_0$.}
\label{fig6}
\end{figure}

Now we consider rolling massive scalar potential $V(\phi)$ given by
\cite{Copeland,Sami},
\begin{equation}
V(\phi) = V_{0}e^{\frac{1}{2}M^{2}\phi^{2}},
\label{eq23}
\end{equation}
In this case, the homogeneous field equation (\ref{eq4}) can be
written as,
\begin{equation}
\frac{\ddot{\phi}}{1 - \dot{\phi}^2}  + M^{2}\phi = 0.
\label{eq24}
\end{equation}
The numerical solutions of this equation is shown in the figure \ref{fig6} by
dotted line with three different colors for three different values of $M$ as 
in case of previous potential, viz., black for 0.1, red for 0.5 and green for 1.
The upper set is for $\phi_0$ = 1 and lower set is for $\phi_0$ = 0.5. The 
field $\phi$ for this potential is oscillatory in nature.
Since $\phi$ is oscillatory just like a simple harmonic motion and its obvious
equation can be obtain by considering $\dot{\phi} << 1$, which is in this case
is,
\begin{equation}
\ddot{\phi} + M^{2}\phi = 0,
\label{eq25}
\end{equation}
and the nature of $\phi$ is the solution of this equation,
\begin{equation}
\phi = \phi_{0}cos(Mt),
\label{eq26}
\end{equation}
where $\phi_{0}$ is a constant, the field amplitude. The solution 
(\ref{eq26}) are presented in the
figure \ref{fig6} for same values of $M$ and $\phi_{0}$ as described above. It 
is seen that the solutions of equations (\ref{eq24}) and (\ref{eq25}) are in 
good agreement, however there is gradual phase shift between them because the 
second solution is due to slow variation of $\phi$. The rate of this phase
shift depends upon the values of $M$ and $\phi_0$, which is slower for the 
smaller values of these two initial field parameters. On the other hand, the 
time period of oscillation of the field depends on these two parameters in the 
opposite way, i. e. the time period increases with decreasing values of them. 
As in the case of the previous potential, in this case also the effect of $M$ 
on the field is more prominent than $\phi_0$. The amplitude of oscillation 
solely depends on the value of $\phi_0$ as it is clear from the figure.    

\begin{figure}[hbt]
\centerline
\centerline{\includegraphics[scale=0.5]{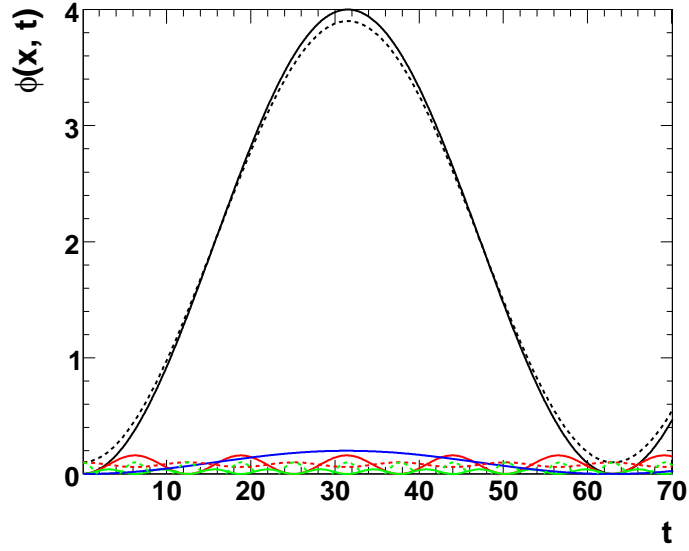}}
\caption{The numerical solutions of the equation (\ref{eq27}) for different
values of $M$, $\phi_0$, $\phi^{\prime}$ and $\phi^{\prime\prime}$. The line
colors black, red and green are for $M$ = 0.1, 0.5 and 1 respectively. Solid
line refer to $\phi_0$ = 0 and dotted to $\phi_0$ = 0.1. All these plots are
drawn for solutions with $\phi^{\prime}$ = 0.01 and $\phi^{\prime\prime}$ = 
0.02. The blue line indicates the solution with $M$ = 0.1, $\phi_0$ = 0,
$\phi^{\prime}$ = 0.002 and $\phi^{\prime\prime}$ = 0.001.}
\label{fig7}
\end{figure}

The inhomogeneous scalar field equation (\ref{eq5}) in this case can be 
written as,
\begin{equation}
\ddot{\phi} = \left(1 - \dot{\phi}^{2} + \phi^{\prime 2}\right)\left(\phi^{\prime\prime} - M^{2}\phi\right).
\label{eq27}
\end{equation}
The numerical solutions of the equation (\ref{eq27}) with $\phi^{\prime} =
0.01$, $\phi^{\prime\prime} = 0.02$ and with three different values of $M$
same as above, and $\phi_0$ = 0, 0.1 are shown in the figure \ref{fig7}. The
three colors corresponds to the respective values of $M$ as mentioned above for
the homogeneous case. Here the solid line indicates the solution for 
$\phi_0$ = 0, whereas the dotted line for the $\phi_0$ = 0.1. To understand the
effect of $\phi^{\prime}$ and  $\phi^{\prime\prime}$ on the inhomogeneous 
field for this potential we consider the solution of the equation (\ref{eq27})
with the set of initial field values as $\phi^{\prime}$ = 0.002, 
$\phi^{\prime\prime}$ = 0.001, $M$ = 0.1 and $\phi_0$ = 0, which is shown by the
blue line in the figure. We have seen that the inhomogeneous field is 
also oscillatory in nature, time period of which depends upon the values of 
$M$ and $\phi_0$ in the same way as in the homogeneous case, however the
small value of $M$ effects significantly on both time period and amplitude of
oscillation. It should be noted that, the amplitude decreases for higher value 
of $\phi_0$ as $M$ decreases and there is a phase shift between oscillation 
patterns of field for different values of $\phi_0$, which decreases with 
decreasing value of $M$. The effect of $\phi^{\prime}$  and 
$\phi^{\prime\prime}$ on the field is only on the amplitude of its oscillation
and we have seen that the amplitude decreases substantially for the lower value
of space variation of the field. So the role of space variation factor is 
significant for this potential to avoid the unwanted large fluctuation or 
oscillation of the field for lower value of $M$. It also implies that in 
reality the space variation of the field should be sufficiently less that 
unity as previously mentioned. 

\begin{figure}[hbt]
\centerline
\centerline{\includegraphics[scale=0.5]{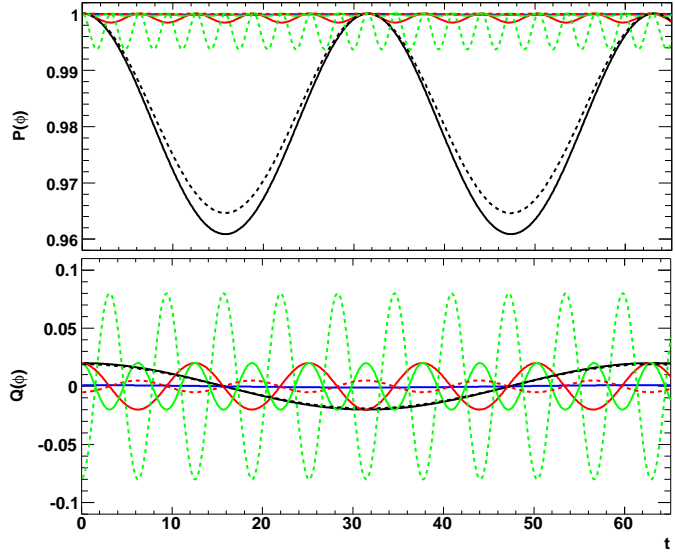}}
\caption{The time variation patterns of $P(\phi)$ and $Q(\phi)$ for the
potential (\ref{eq23}) corresponding to the solutions of equation 
(\ref{eq27}) shown in the figure \ref{fig7}.}
\label{fig8}
\end{figure}

\begin{figure}[hbt]
\centerline
\centerline{\includegraphics[width = 8cm, height = 8cm]{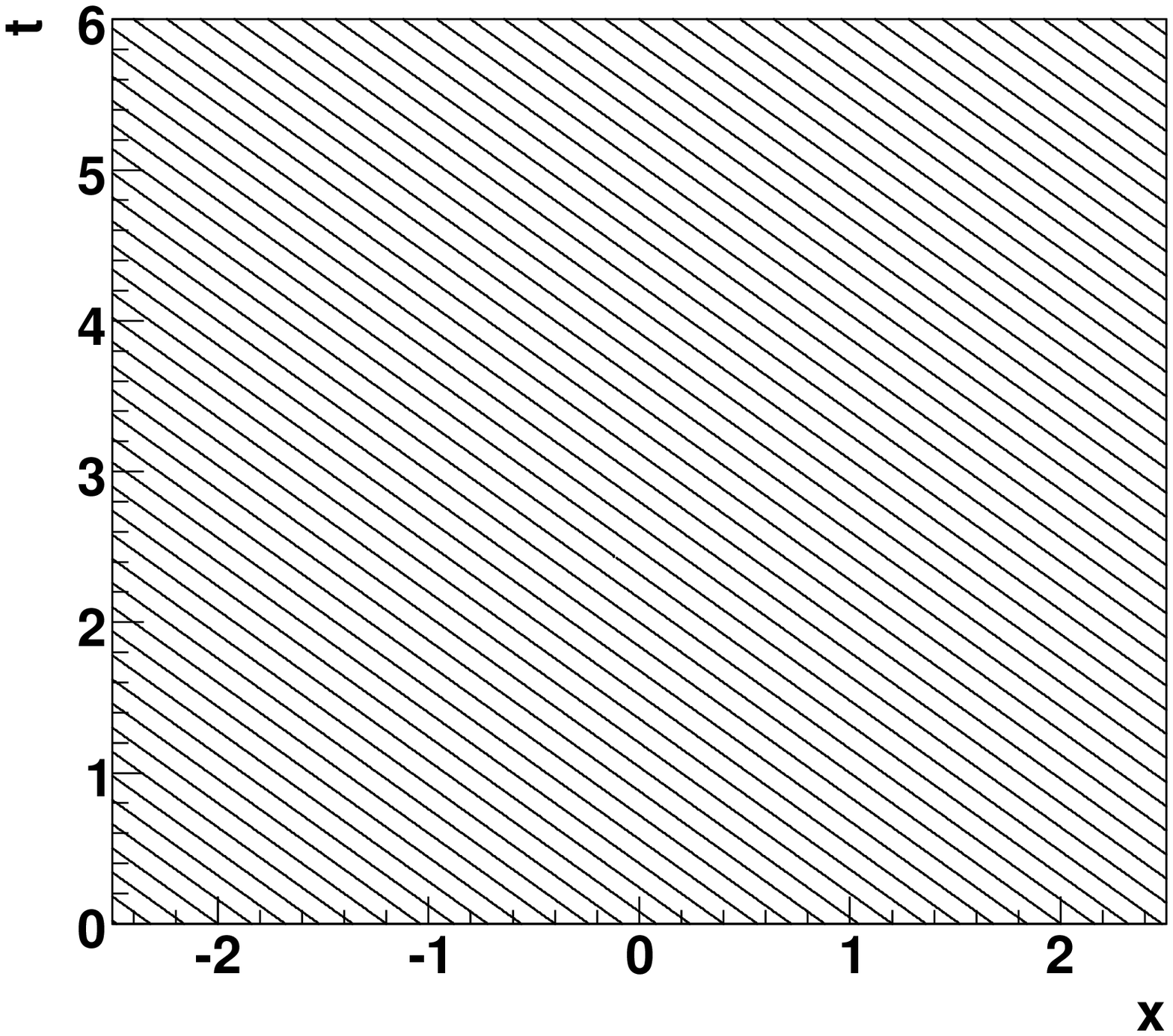}
\includegraphics[width = 8cm, height = 8cm]{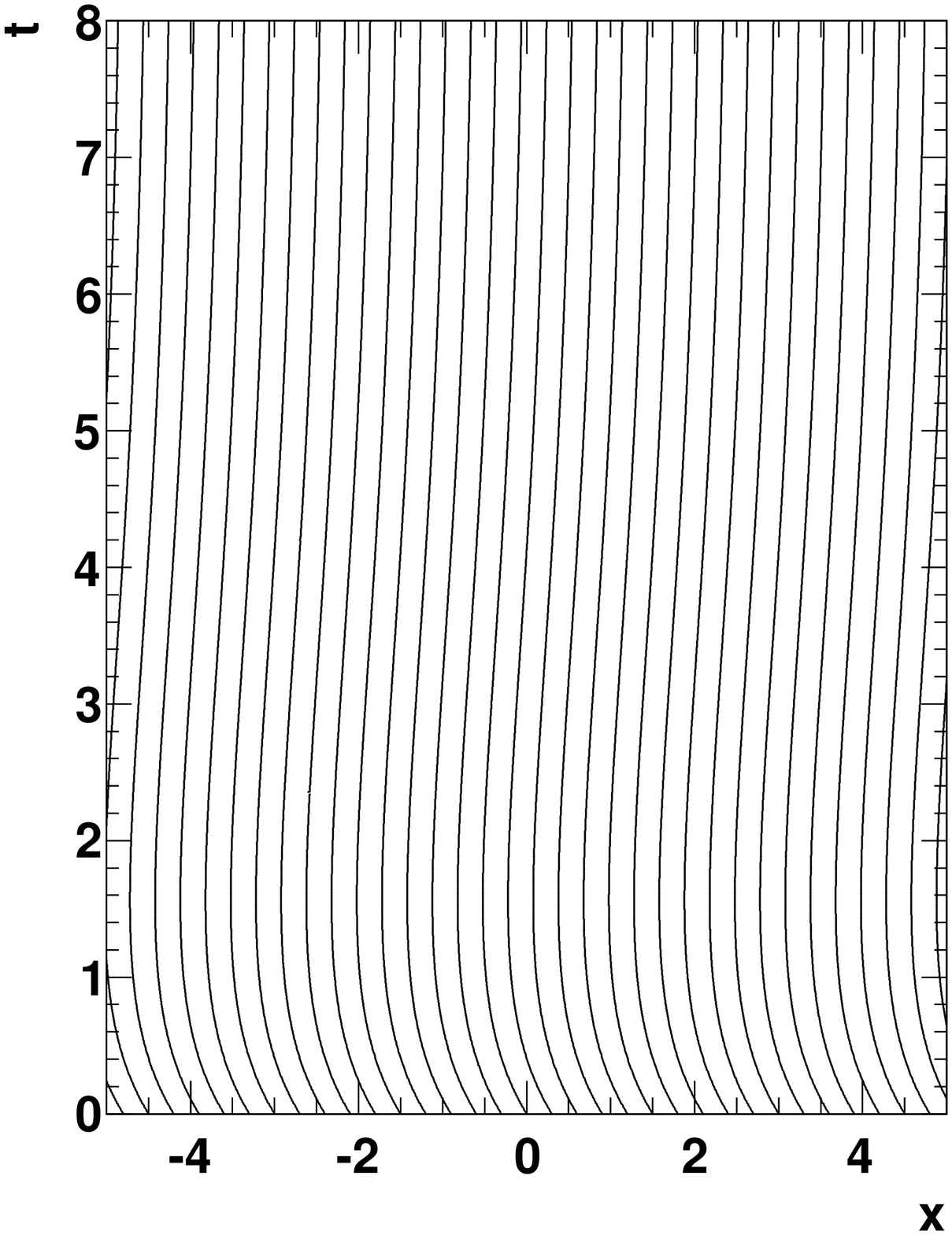}}
\caption{The particle trajectories in 1 + 1 Minkowski space obtain from 
equation (\ref{eq16}) (left) and in FRW expanding universe obtain from
equation (\ref{eq5d}) (right) with potential (\ref{eq23}) and 
$\phi_{i}(q) = exp(-q^{2})$. The figure in the right panel is obtained from the
numerical solutions of equations (\ref{eq4a}) and (\ref{eq4b}) of FRW expanding
universe with the given potential for one spatial dimension taking 
constants $k= 8\pi G$, $V_0$, $M$ as unity and the initial field value 
$\phi_0$ = 0.1.}
\label{fig9}
\end{figure}

The time variations of $P(\phi)$ and $Q(\phi)$ corresponding to above solutions
of equation (\ref{eq27}) are shown in the figure \ref{fig8}.
The values of $P(\phi)$ oscillates with time near unity from below with 
different amplitudes corresponding to the situations described above for the
solutions of equation (\ref{eq27}). For the value of $M$ = 0.1 the amplitude of
oscillation of values of $P(\phi)$ is noticeably away from unity in comparison 
to other higher values of $M$ if we consider the value of $\phi^{\prime}$ = 
0.01 and $\phi^{\prime\prime}$ = 0.02. So for further lower values of $M$ the 
amplitude of oscillation should move far away from unity. On the other hand 
if we consider the value $\phi^{\prime}$ = 0.002 and $\phi^{\prime\prime}$ = 
0.001 for the same value of $M$ (i.e. 0.1), the oscillation of the values of
$P(\phi)$ is negligible and  its values are almost unity for all times. But 
intuitively enough, we may argue that the value of space variation of the 
field can not be very small beyond some limit as that will nullify the 
inhomogeneity condition of the field and similarly the value $M$ also can not 
be very small because then field will not be sustainable one for very large 
amplitude of oscillation. Hence for the real situation the value of $M$ and 
the space variation of the field should be such that oscillation of the field
must be negligible. Under such considerations and for our present values of 
$M$ we may take on average the value of $P(\phi)$ as unity for all times. For 
the initial field configuration $\phi_{i}(q) = exp(-q^{2})$ as in the previous 
cases and $P(\phi) = 1$, the characteristic curves for equation (\ref{eq16}) 
are shown in the left panel of the figure \ref{fig9}.
It is clear from the panel of the figure that there are no caustics
and there are no multivalued regions in the field configuration.
This happens due to the almost steady nature of the field (because the average
value of the field with time is a constant as clear from our argument together
with the figure \ref{fig7}) and therefore the evolution pattern of the field 
for all times in this case will be same as in the previous case for $t = 0$, 
i. e. same as the initial configuration of the field. 
As expansion works against caustics formation, we should not encounter 
caustics if expansion of universe is incorporated. Indeed, we have
numerically checked that there are no caustics in the field profile
in the expanding universe in this case, see the right panel of the 
figure \ref{fig9}.
\subsection{Inverse power-law potentials}
Finally we consider the inverse power-law potential of the form,
given by \cite{Copeland,Sami},
\begin{equation}
V(\phi) = V_{0}\phi^{-n},~~0<n<2.
 \label{eq28}
\end{equation}
The late time accelerated expansion of the universe corresponds to
$0\le n \le 2$ \cite{Abramo,Copeland}, and accordingly we restrict
the value of the exponent $n$ within this range for our work.

\begin{figure}[hbt]
\centerline
\centerline{\includegraphics[scale=0.25]{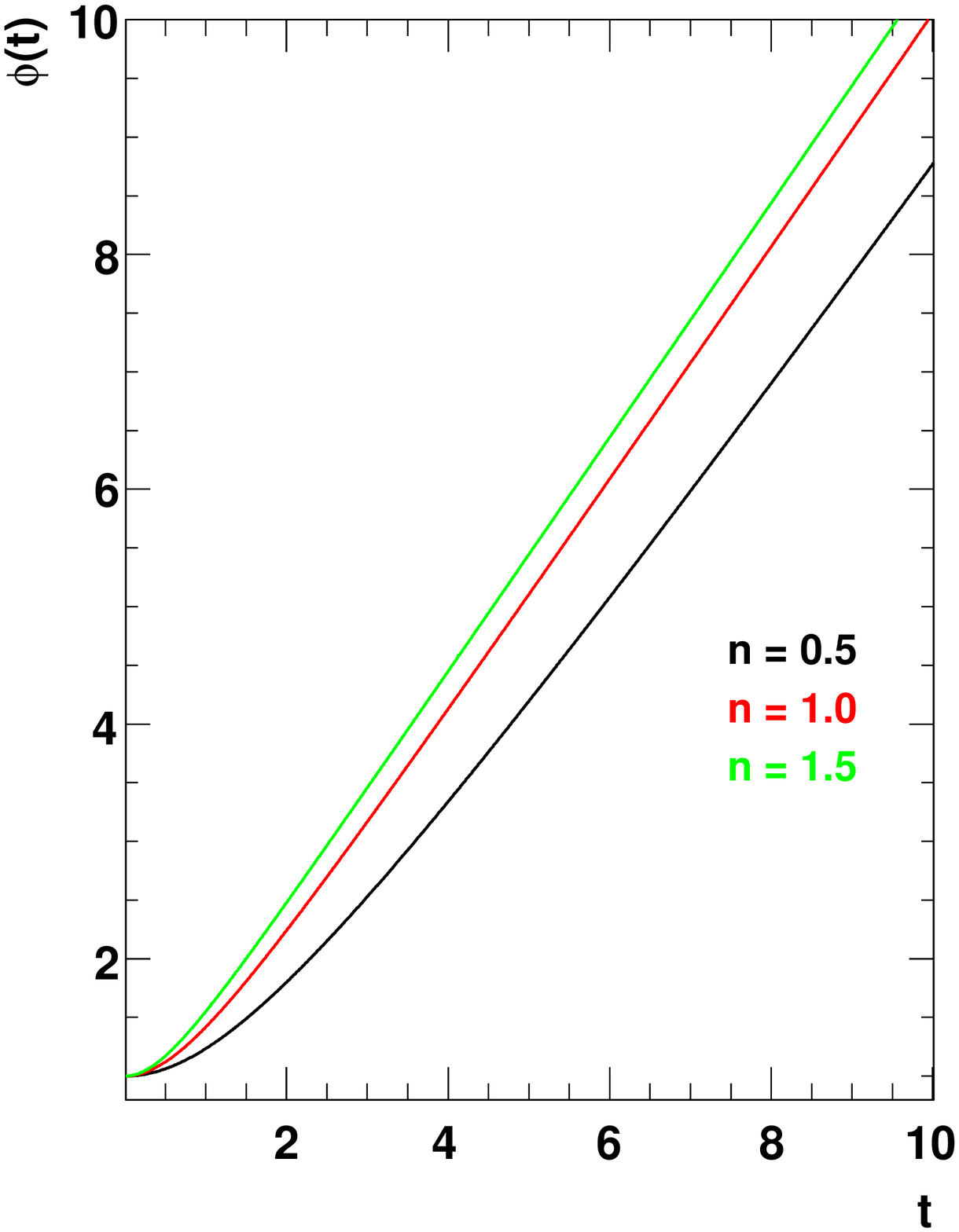}
\includegraphics[scale=0.25]{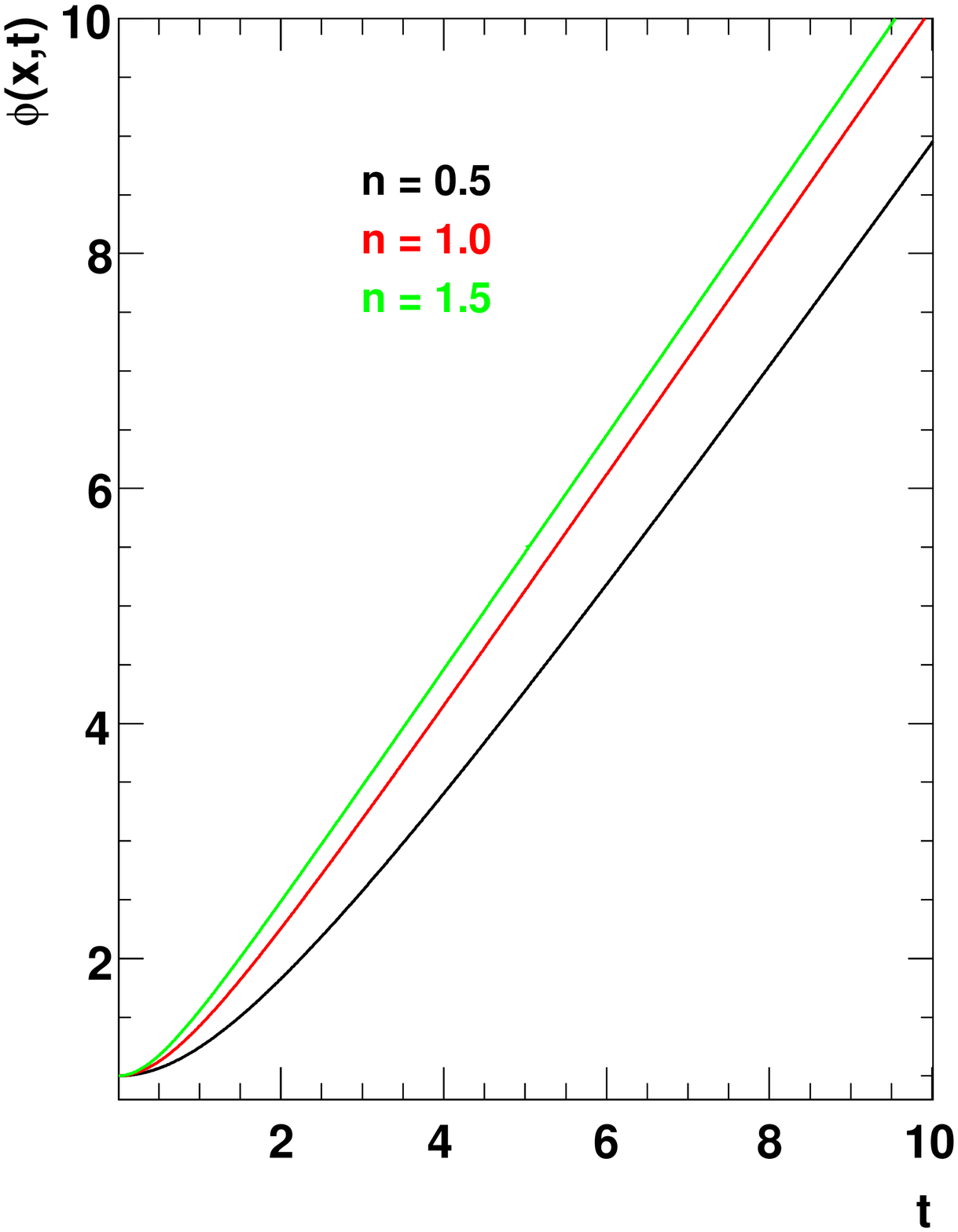}
\includegraphics[scale=0.25]{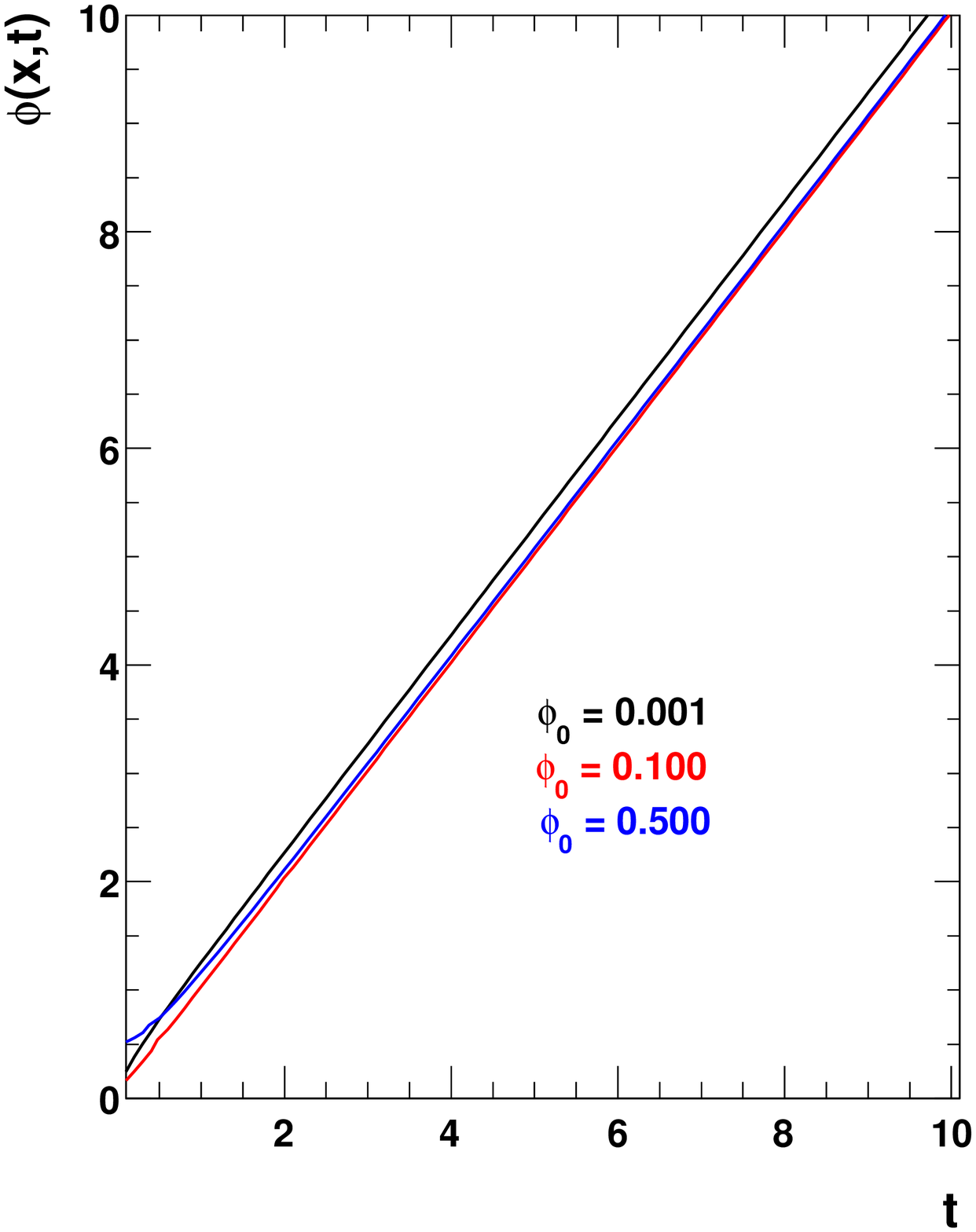}}
\caption{The numerical solutions of the homogeneous field equation
(\ref{eq29}) (left) and the inhomogeneous field equation (\ref{eq30})
(middle) with $\phi_0$ = 1. The difference between the homogeneous and 
inhomogeneous solutions is insignificant. The right panel shows the three 
different solutions of the inhomogeneous equation for three different values
of $\phi_0$ and for $n$ = 1. This panel clearly verifies that the time 
evolution of the field is almost independent of the initial value of the field 
for the inverse power-law potentials.}
\label{fig10}
\end{figure}

\begin{figure}[hbt]
\centerline
\centerline{\includegraphics[scale=0.25]{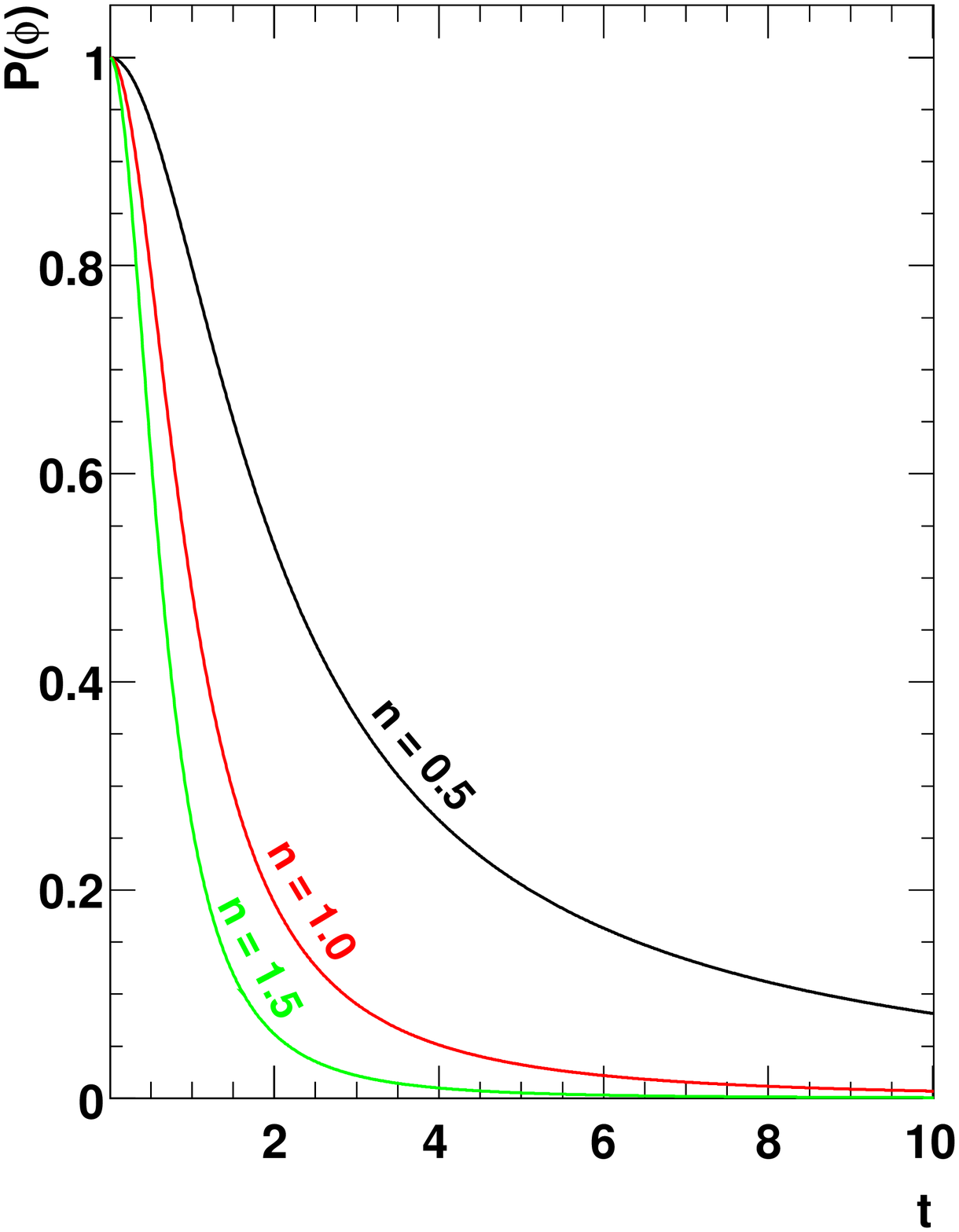}
\includegraphics[scale=0.25]{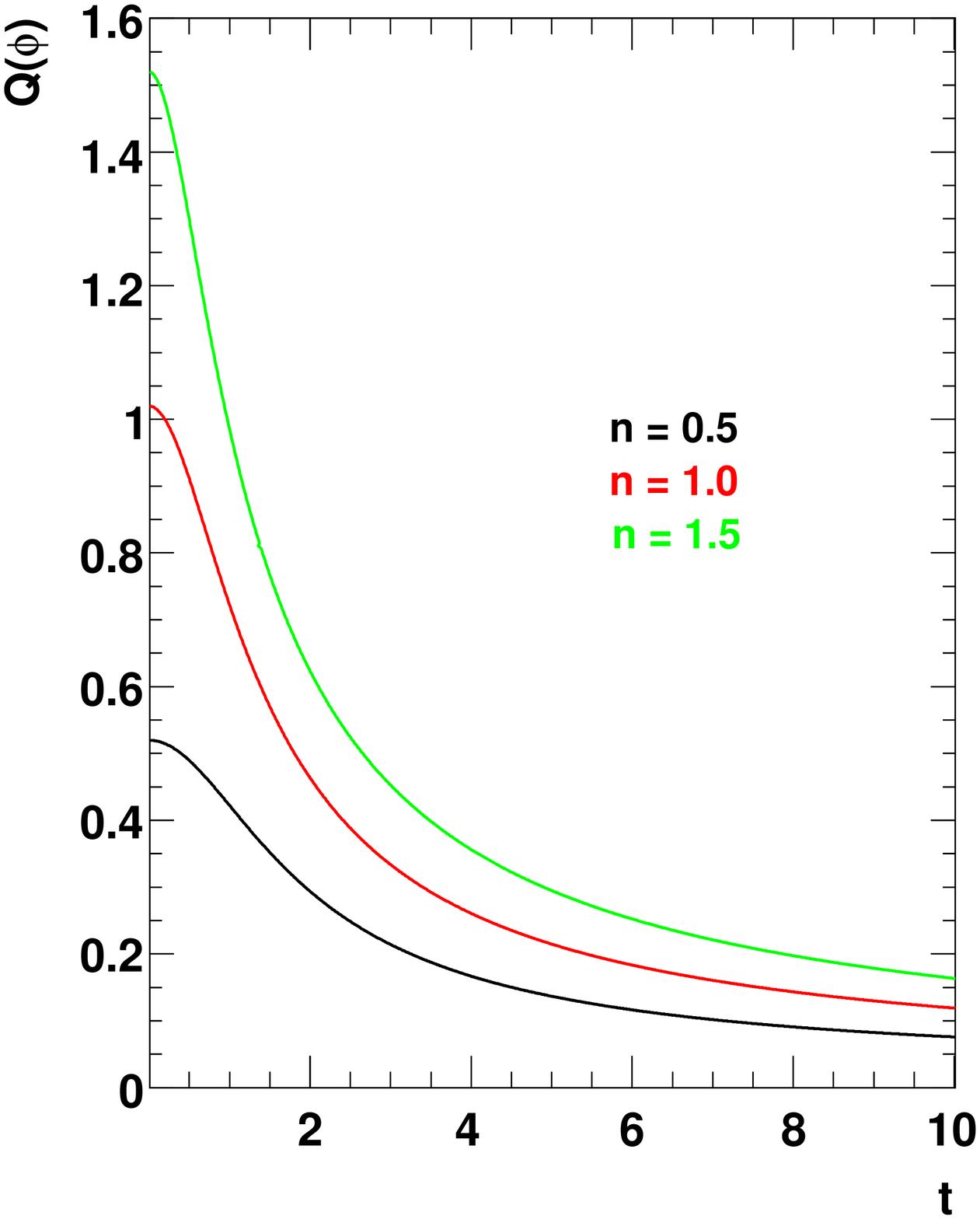}
\includegraphics[scale=0.5]{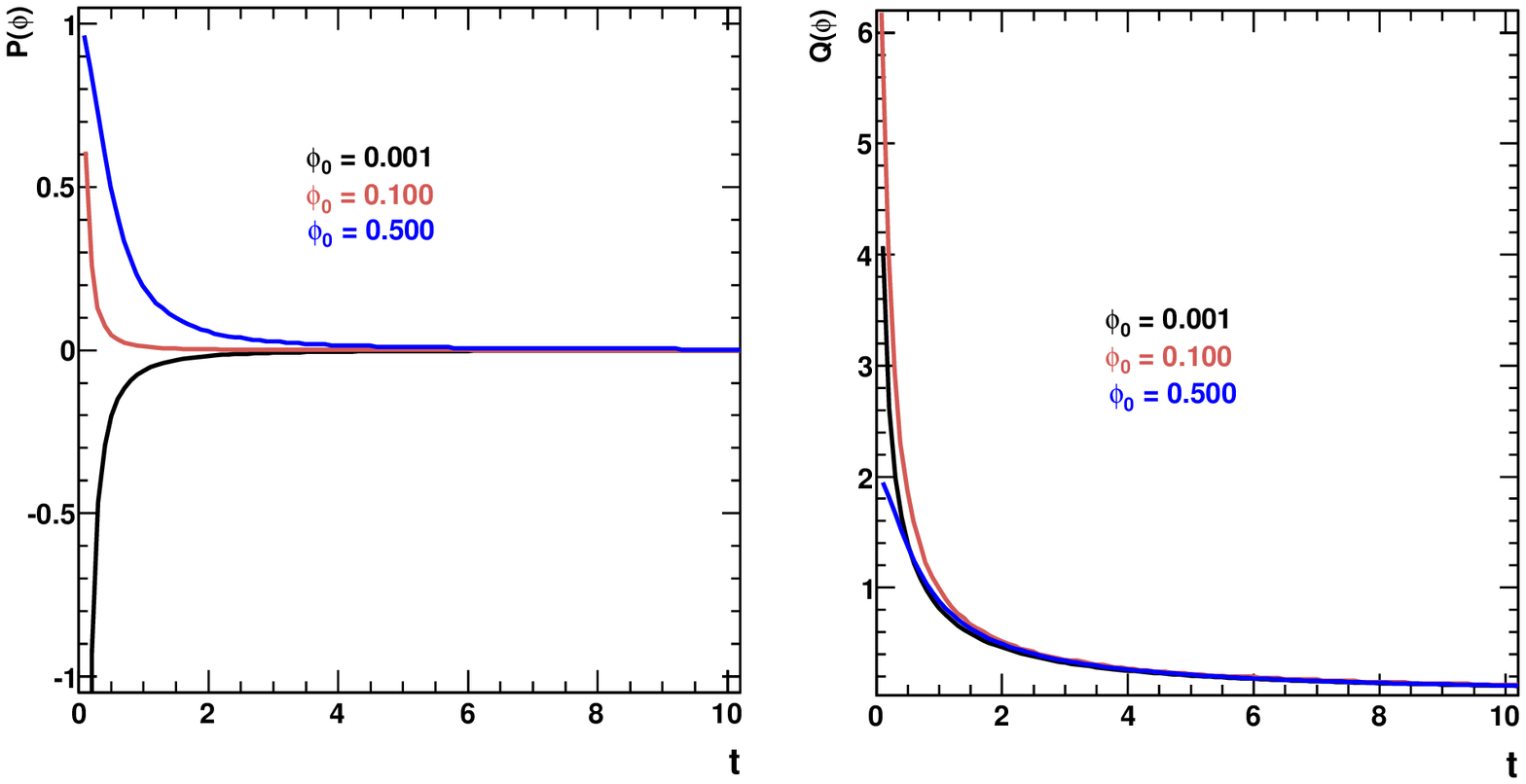}
}
\caption{The time variation patterns of $P(\phi)$ and $Q(\phi)$ corresponding 
to the solutions shown in the middle and the right panels of the 
figure \ref{fig10}.}
\label{fig11}
\end{figure}

We shall first restrict our attention to 1+1 dimensional Minkowski space time. 
In this case, the homogeneous field equation (\ref{eq4}) is given
by,
\begin{equation}
\frac{\ddot{\phi}}{1 - \dot{\phi}^2}  - \frac{n}{\phi} = 0.
\label{eq29}
\end{equation}
The numerical solutions of this equation is shown in the left panel
of the figure \ref{fig10} for $n = 0.5,\;1.0\;\mbox{and}\; 1.5$ with 
$\phi_0$ = 1. The difference of the solutions decreases for the higher values 
of $n$ as it is clear from the figure.
The inhomogeneous scalar field equation (\ref{eq5}) in this case will take
the form,
\begin{equation}
\ddot{\phi} =
\left(1 - \dot{\phi}^{2} + \phi^{\prime 2}\right)\left(\phi^{\prime\prime} + \frac{n}{\phi}\right).
\label{eq30}
\end{equation}
The numerical solutions of the equation (\ref{eq30}) with $\phi^{\prime} =
0.01$, $\phi^{\prime\prime} = 0.02$ and for $n = 0.5,\;1.0,\;1.5$ with 
$\phi_0$ = 1 are shown in the middle panel of the figure \ref{fig10}. It 
should be noted that there is no significant difference between the solutions 
of homogeneous and inhomogeneous field equations for this inverse power-law 
potential, because the time variation of the field is more prominent than the 
space variation we have considered, which indeed should be for a mass free 
space. Moreover the field evolution pattern is almost independent of the
initial condition of the field. However it has little impact on the magnitude 
of the field as time passage in a arbitrary manner as it is seen from the 
right panel of this figure. So the value of $\phi_0$ would be immaterial for
the testing of the caustic formation in the field for the inverse power-law
potentials. However we take care of the effect of $\phi_0$ on the variation
$P(\phi)$ to see the way of rolling of its values with time which will be 
clear from the following discussion.     
The time variations of $P(\phi)$ and $Q(\phi)$ corresponding to
the solutions of $\phi(x, t)$ as shown in the middle and the right panel
of the figure \ref{fig10} are shown in the figure \ref{fig11}.
\begin{figure}[hbt]
\centerline
\centerline{\includegraphics[width = 8cm, height = 8cm]{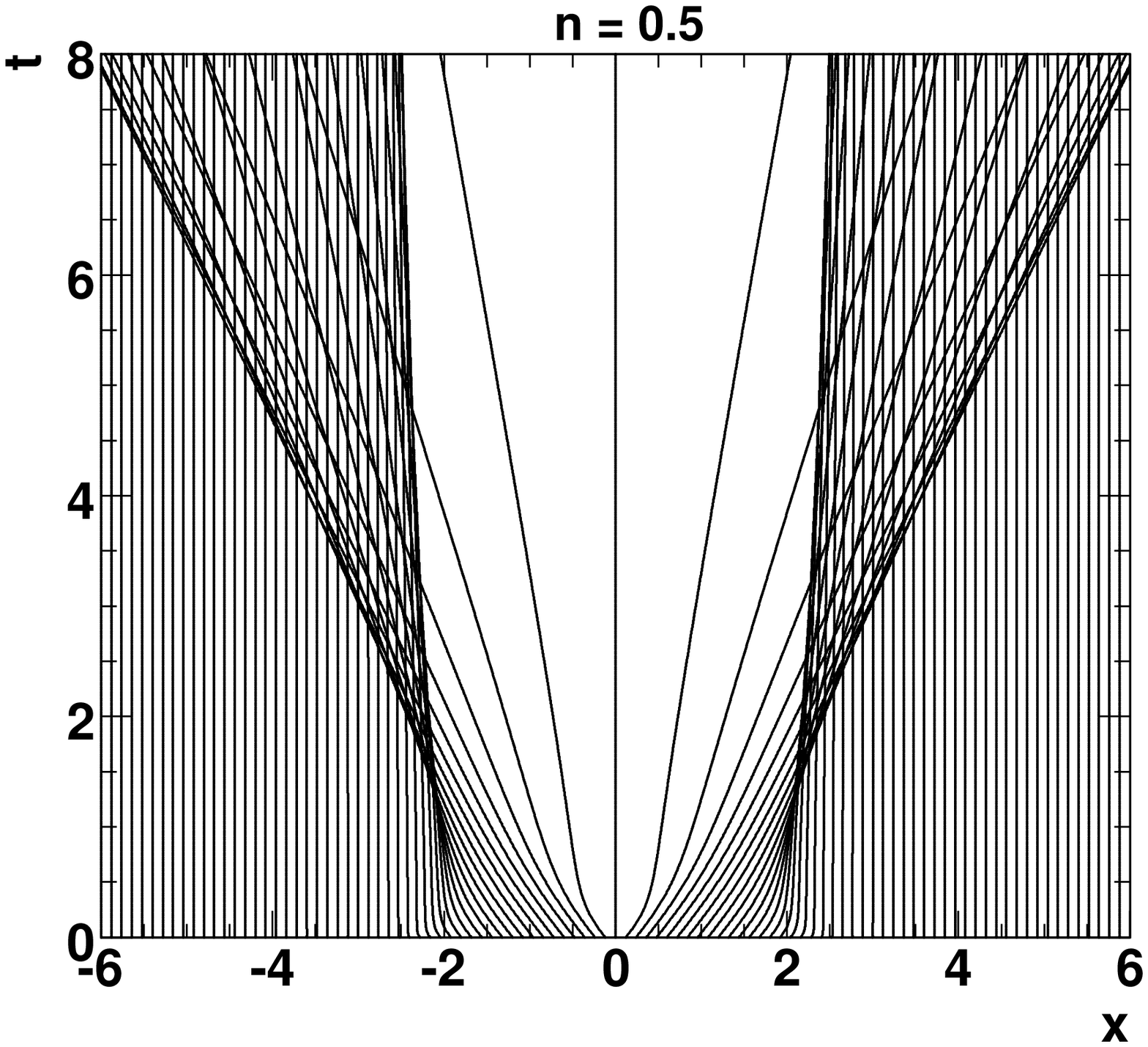}
\includegraphics[width = 8cm, height = 8cm]{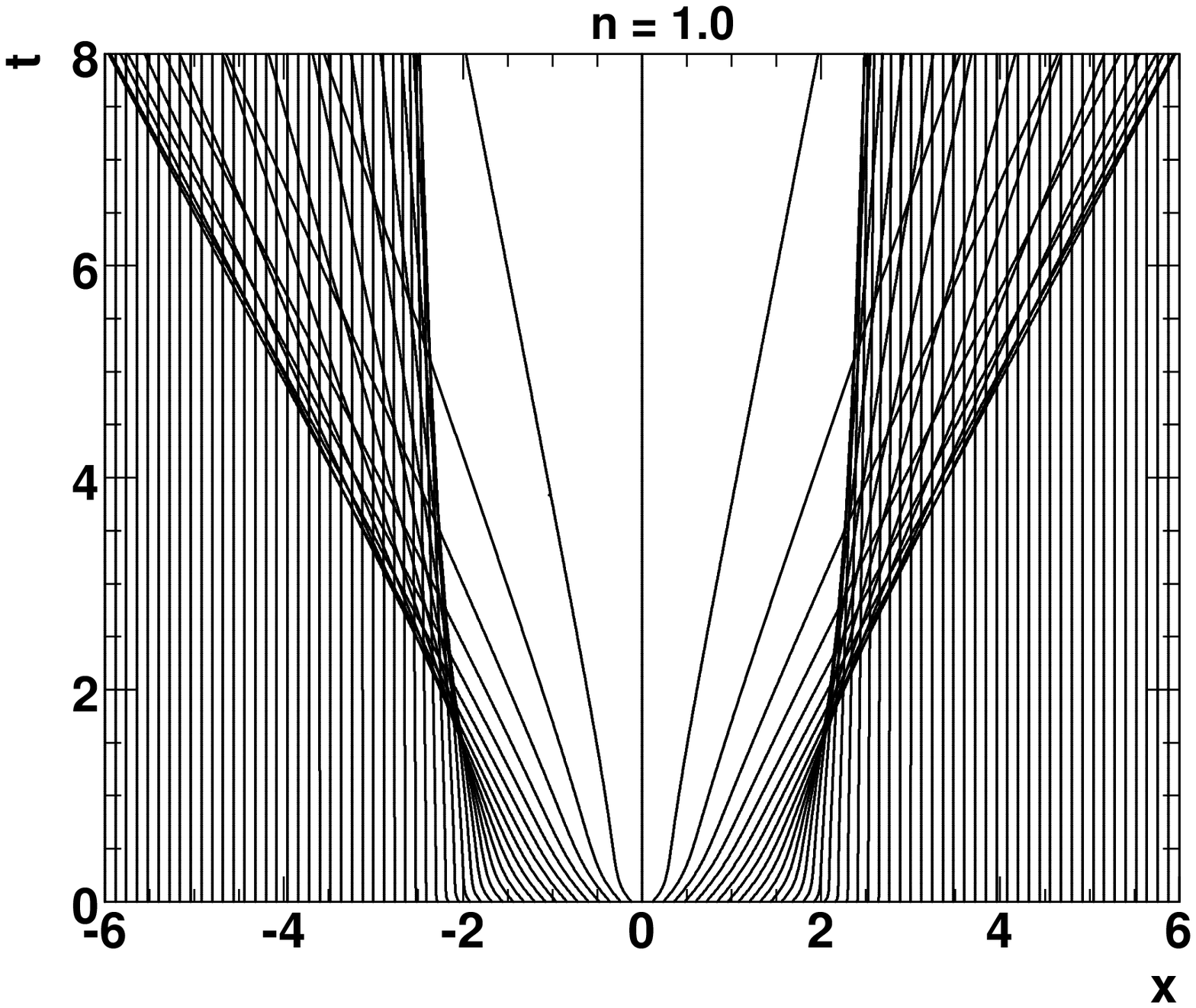}
\includegraphics[width = 8cm, height = 8cm]{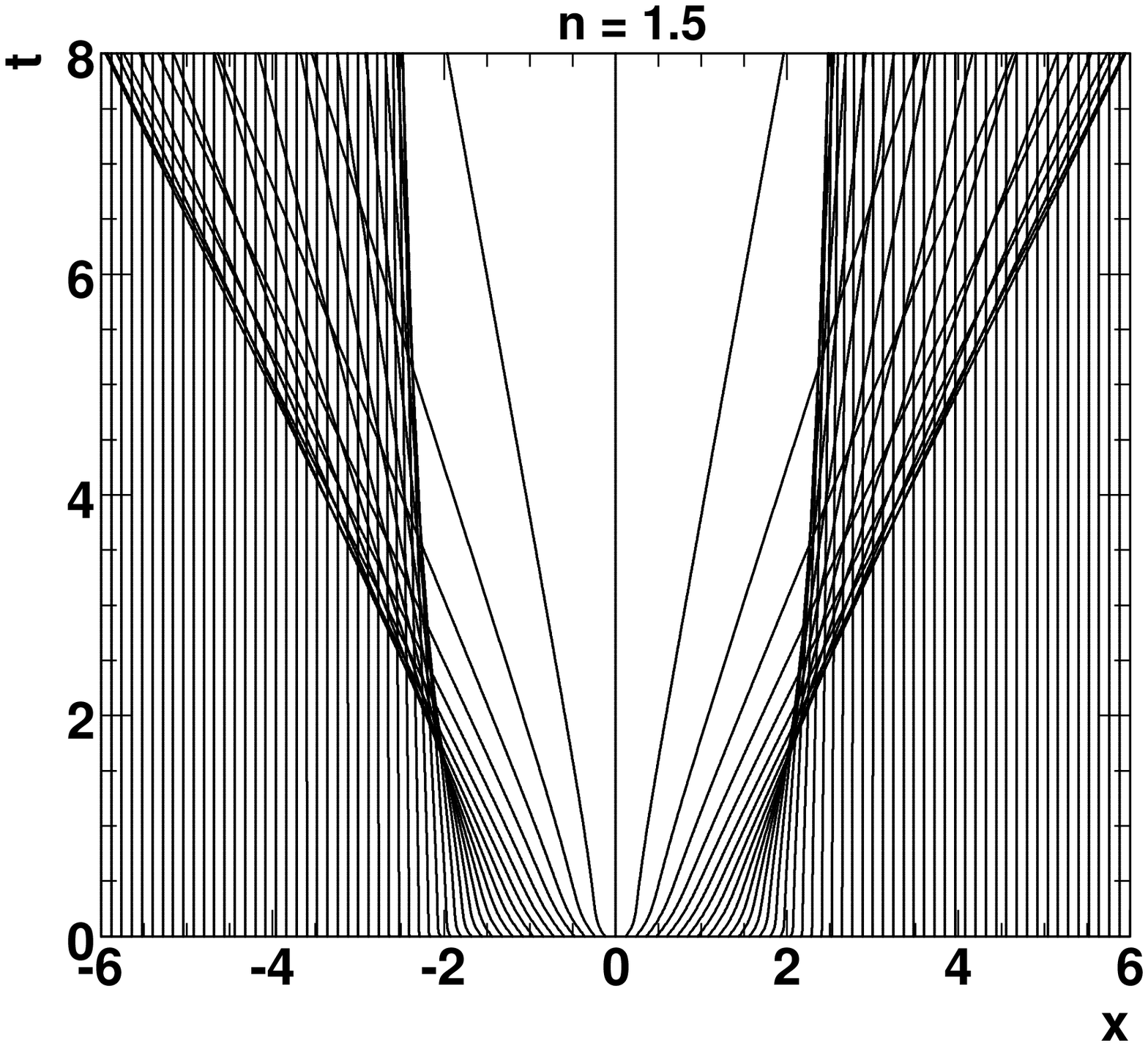}
}
\caption{The particle trajectories in 1 + 1 Minkowski space obtain from 
equation (\ref{eq16}) with potential (\ref{eq28}) and 
$\phi_{i}(q) = exp(-q^{2})$ corresponding to the solutions of the equation 
(\ref{eq30}) with $\phi_0$ = 1 as shown in the middle panel of the 
figure \ref{fig10}. Caustics and multi-valued regions are clearly formed in
the space.}
\label{fig12}
\end{figure}

From the figure \ref{fig11} we observed that values $P(\phi)$ for all $n$
rapidly falls to zero from above if the initial value of the field $\phi_0$
is not very small. It is interesting to note that, when the value of $\phi_0$
is very small, $P(\phi)$ falls rapidly to zero from below as indicated by the 
third panel of the figure \ref{fig11}. Notwithstanding, for all cases
the value of $P(\phi)$ rolls to zero very fast (rate is more for lower value
of $\phi_0$)  as time passage and hence considering the initial field 
configuration as in the previous cases, the plots of the characteristic curves 
for equation (\ref{eq16}) are shown in the figure \ref{fig12} for 
$n = 0.5,\;1.0\; \mbox{and}\;1.5$ with $\phi_0$ = 1.

The figure \ref{fig12} vividly shows the formation of caustics and multi-valued
regions in this case. It should be noted that the regions of caustics are not
steady but slowly moves with time as the exponent $n$ of the potential
increases. Due to pattern of variation of $P(\phi)$, we  get the
similar pattern of variation of field configuration at different times as in 
the case of exponentially decreasing rolling massive potential as shown in the
figure \ref{fig5}.

\begin{figure}[hbt]
\centerline
\centerline{\includegraphics[width = 8cm, height = 8cm]{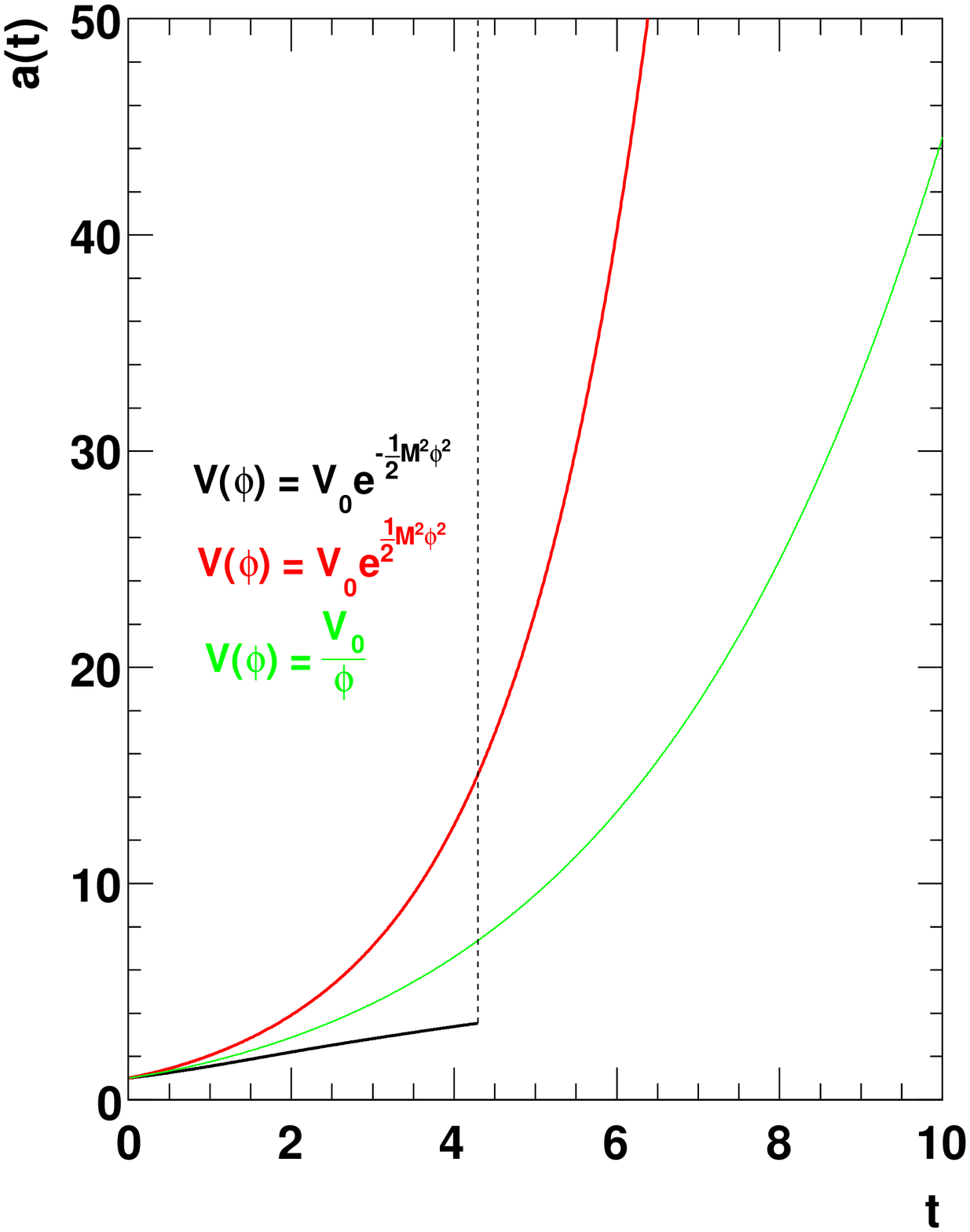}
\includegraphics[width = 8cm, height = 8cm]{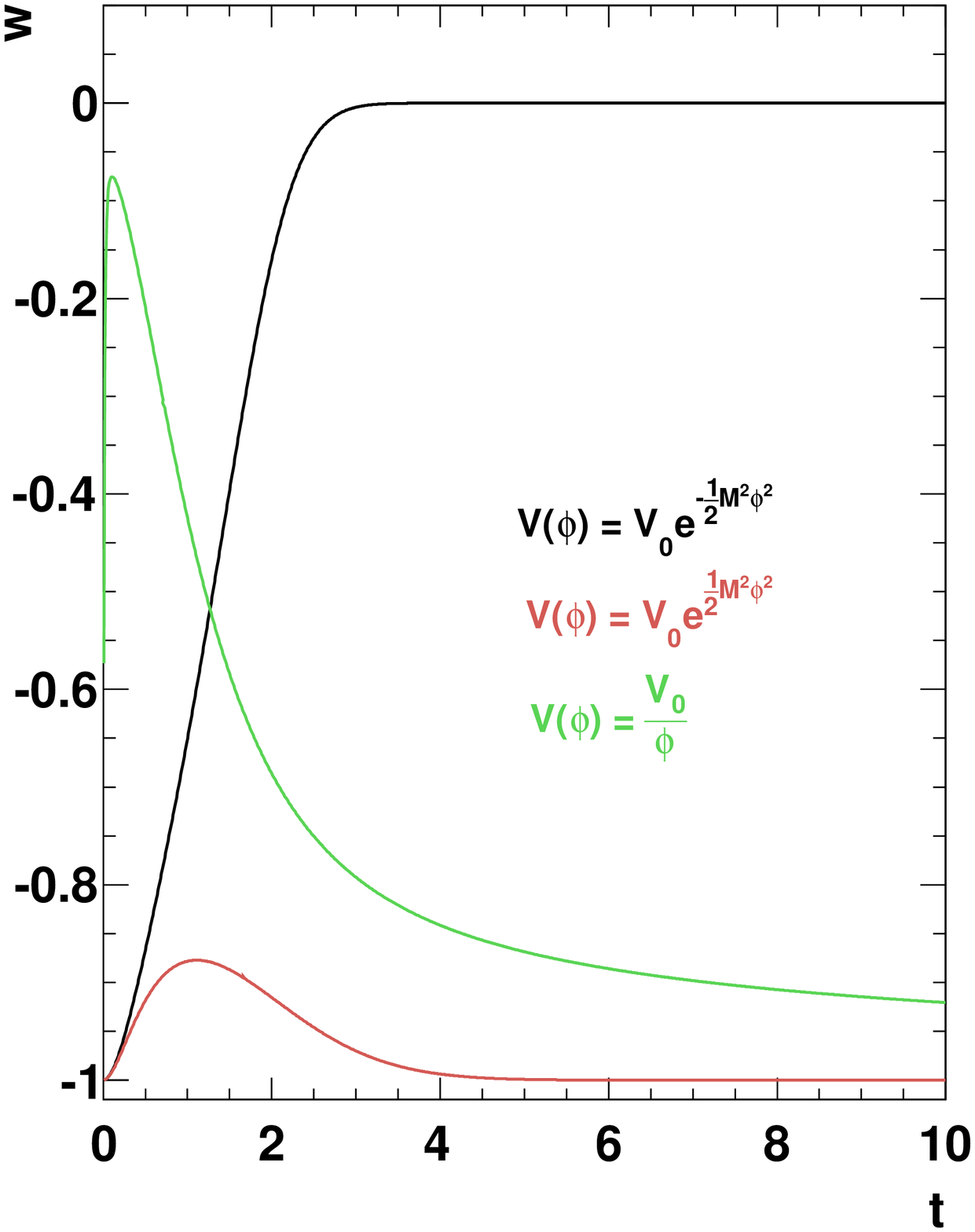}
}
\caption{Numerical solutions of equation (\ref{eq4b}) for the three different
field potentials (left). The dotted line in the figure indicates the infinite
value. The time variations of the equation state parameter $w$ for all three 
potentials (right). These plots are obtained by taking constants $k= 8\pi G$, 
$V_0$, $M$, $n$ as unity and the initial field value $\phi_0$ = 0.1.}
\label{fig13}
\end{figure}

As noted earlier, expansion works against caustic formation. The effect
might become dramatic in case of inverse power-law potential under 
consideration with dark energy as late attractor. Thus the observed caustic 
formation in case of potential (\ref{eq28}) may quite be the artifact of 
Minkowski space time{\footnote{We thank A. Starobinsky for bringing this point 
to our notice}}. In order to reach the final conclusion in this regard, it is 
essential to incorporate the expansion of universe. The numerical solutions of 
evolution equations (\ref{eq4a}) and (\ref{eq4b}) are depicted in the 
figure \ref{fig13} for all potentials, which are obtained by taking constants 
$k= 8\pi G$, $V_0$, $M$, $n$ as unity and the initial field value 
$\phi_0$ = 0.1. The left panel provide us the 
evolution history of the scale factor $a(t)$ and right panel shows the 
variation of the equation state parameter $w$ of the field with time for all 
three potentials. It is clearly observed that, the time evolution of the
scale factor is faster in the case of exponentially increasing massive
potential than the inverse power-law potential. On the other hand in case
of exponentially decreasing massive potential the scale factor becomes 
infinity after some initial period. The field is rolling faster to
attain steady values in the case of the exponentially increasing and decreasing
field potentials. Whereas in the case of the inverse power-law potential
($0<n<2$, the plot is shown for $n = 1$) the field is evolving very slowly
with time after some very brief initial period during which it evolves
relatively fast. Though both the models (i. e. exponentially increasing and
inverse power-law) can account for late time acceleration, the rolling massive 
scalar requires enormous fine tuning in order to be consistent with observation.

\begin{figure}[hbt]
\centerline
\centerline{\includegraphics[width = 8cm, height = 8cm]{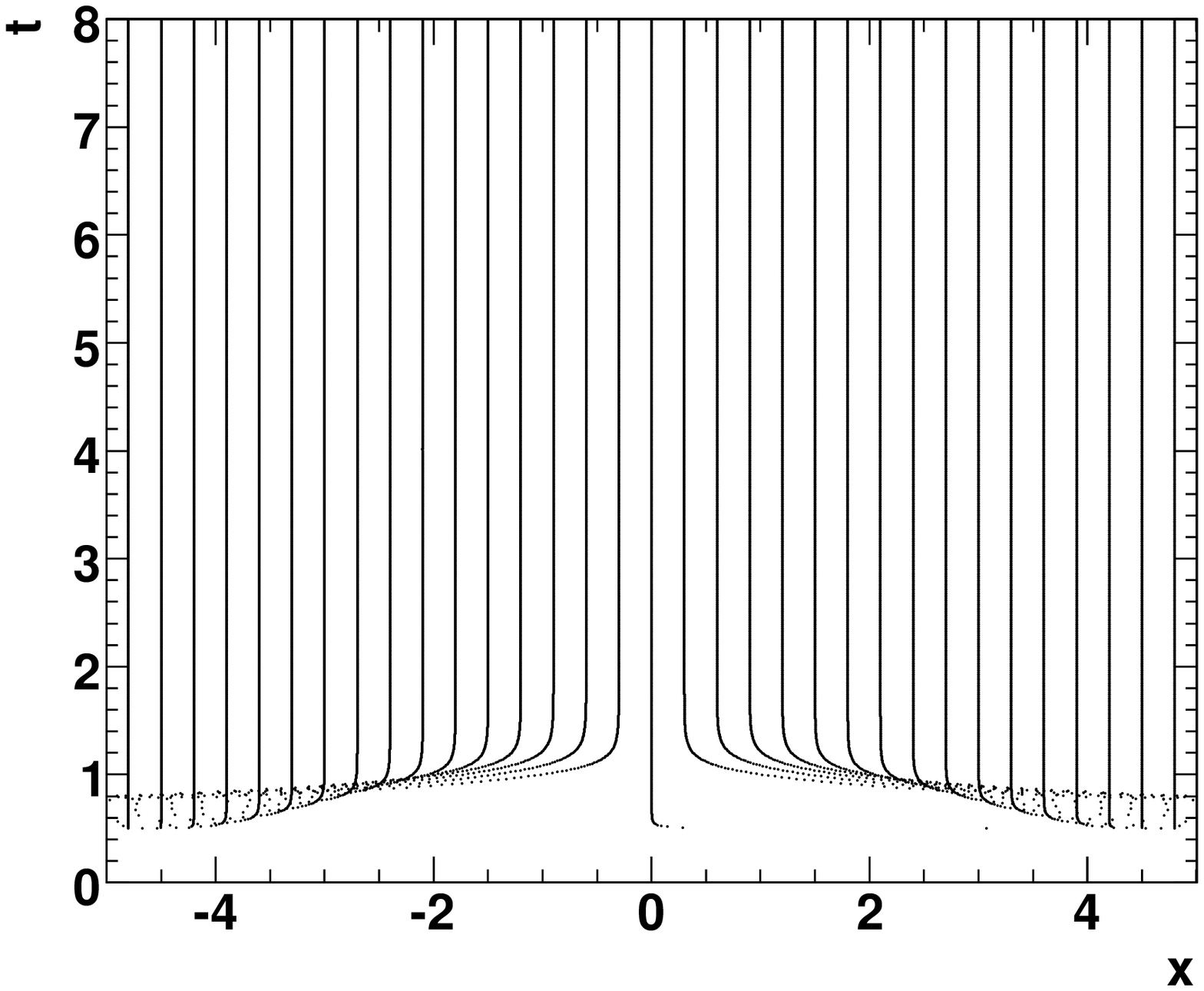}
\includegraphics[width = 8cm, height = 8cm]{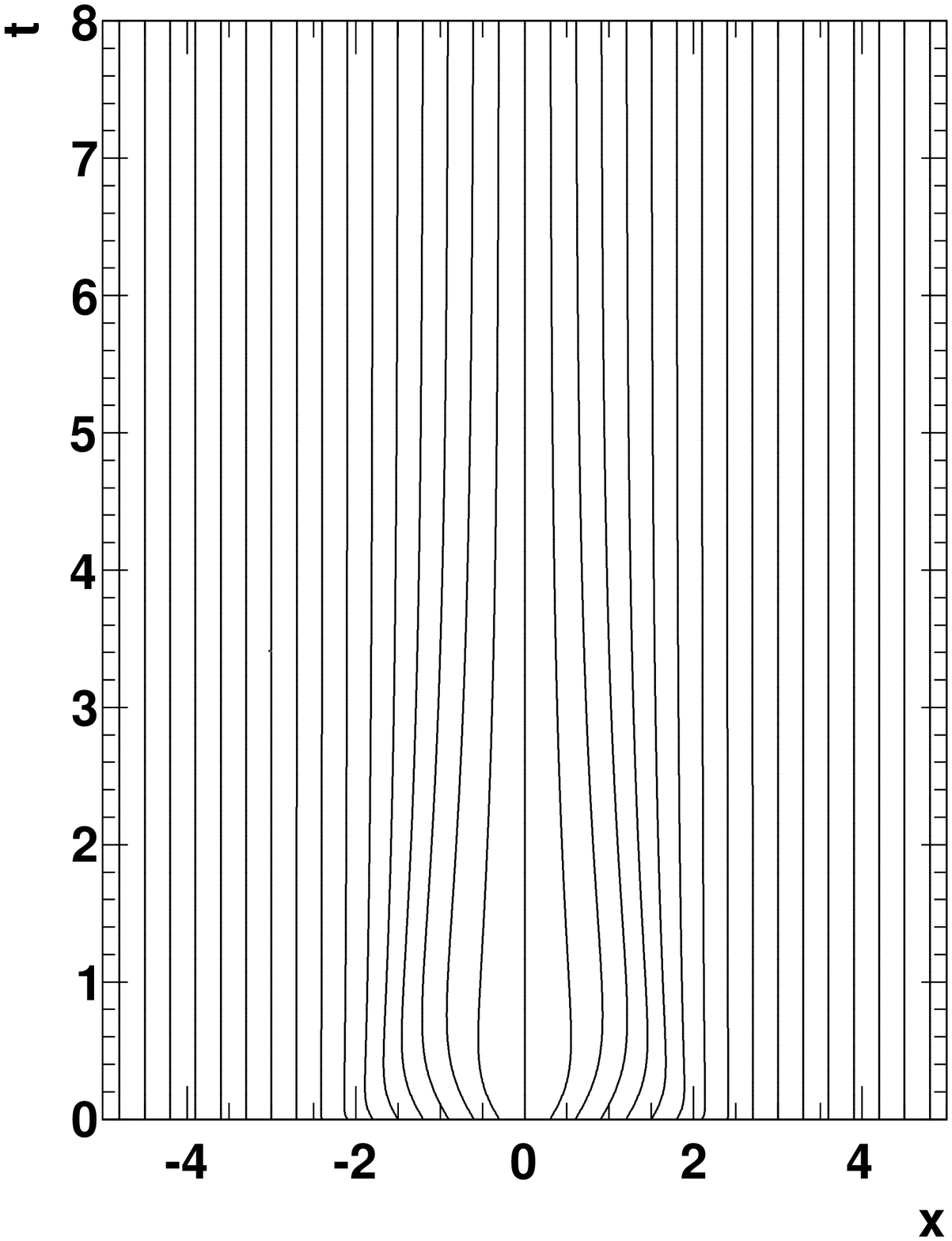}}
 \caption{The particle trajectories in FRW expanding universe obtain from 
equation (\ref{eq5d}) assuming all time dark energy effect, i. e. with the
scale factor $a\sim t^{2/3(1+w)}$ for $w$ = - 0.9 (left) and using the 
numerical data of solutions of equations (18) and (19) (right) for the inverse 
power-law field potential (\ref{eq28}) and the profile $\phi_{i}(q) = 
exp(-q^{2})$. Other assumptions for the right panel of this this figure are 
same as the right panel of the figure \ref{fig9}. We put both left and right 
panels in this figure to focus on the late time evolution of the field as well 
as to show that at late times the inverse power-law potentials have exactly 
same effect on the scalar field as the dark energy has on it.} \label{fig14}
\end{figure}

Since dark energy is late time attractor in the present case, we first for 
simplicity and consistency with the full solution of $a(t)$ as mentioned
above, assumed that $a\sim t^{2/3(1+w)}$, incorporating all time dark energy
effect in the scale factor. Now for the case of the scalar field in expanding 
universe with the inverse power-law potentials under the above mentioned 
assumptions and with this value of the scale factor for $w$ = - 0.9, we obtain 
the characteristic curves of the field given by the equation (\ref{eq5d}) look 
like as shown in the left panel of the figure \ref{fig14}. It is 
interesting to note that there are no caustics and multivalued regions in the 
field profile of dark energy dominated universe in contrast to the 
situation of 1 + 1 dimensional Minkowski space. Finally at this stage we 
intend to incorporate the dynamics exactly by taking the numerical solutions of
equations (\ref{eq4a}) and (\ref{eq4b}). The numerical data of the scale 
factor $a(t)$ (see, figure \ref{fig13}) can then be used to solve the equation 
(\ref{eq5d}) to see the formation of caustics in the field profile for the 
given potential. The result of solution of this equation (\ref{eq5d}), i. e.
the characteristic curves obtained by using the numerical data of the 
scale factor is shown in the right panel of the figure \ref{fig14} by
considering the same initial profile of the field as in the previous cases. 
Both of the panel of this figure is for $n$ = 1. It is apparent from the 
figure \ref{fig14} in the right panel that at late times ($>4$), the field 
profile based upon the exact simulation is same as the one (left panel of 
figure \ref{fig14}) that is obtained by assuming dark energy domination at all 
times. This shows that at late times the inverse power-law potentials have 
exactly same effect on the scalar field in the expanding universe as the dark
energy has on it. Since the dark energy is a
late time effect, so the disagreement of the two panels of this figure
during inital evolution is obvious and does not have any significance for us.     


In view of the aforesaid, we consider the models based upon the inverse 
power-law potentials more viable than the one with rolling massive scalar 
field to explain the present accelerated expansion of our universe.

\section{Conclusion}
In this paper, we have examined the phenomenon of caustic formation
in tachyon system with two generic classes of potentials. We
presented analytical estimates supported by detailed numerical
simulation. We show that the time variation of the scalar field with
exponentially decreasing  potential,
$V=V_0e^{-\frac{1}{2}M^2\phi^2}$, becomes gradually important as the
time elapses for both the homogeneous and the inhomogeneous field
configurations. There are multi-valued regions and regions of likely
to be caustic for this potential in Minkowski space time as well as in
FRW expanding universe which broadly agrees with the analysis of
Ref.~\cite{staro}. However in the case of the expanding universe, the
caustic formations is more definite than the case of 1 + 1
dimensional Minkowski space. It is clearly seen from figure \ref{fig4} that 
expansion
works against caustic formation: it dilutes the effect of caustics but can
not render the situation caustic free in the present case which is generally 
true for a tachyon potential that decays faster than $1/\phi^2$ at infinity.

The field exhibits oscillatory behavior
for exponentially increasing rolling massive scalar field potential,
$V=V_0e^{\frac{1}{2}M^2\phi^2}$. In the case of inhomogeneous
field, the magnitude of time oscillations of the field is very small
and hence the field can be considered as almost steady for a sufficiently
tuned values of initial parameters. The field
with exponentially increasing rolling massive potential is free from
caustic formation in both cases of 1 + 1 dimensional Minkowski space
and expanding universe, see Fig. 9. It is interesting to note that
this potential can give rise to late time acceleration provided that
we fine tune the energy scale $V_0$ in the potential appropriately.

For inverse power-law potentials, $V(\phi)=V_0/\phi^n,~0<n<2$, dark
energy is a late time attractor of dynamics and in this case, we did
expect the system to be free from caustics. Our analysis shows that
the time variation of the homogeneous and inhomogeneous fields are
almost similar in this case, because the space variation of the
field is insignificant in comparison to time variation. Fig. 12 shows
that caustics are formed with the multi-valued regions beyond them
in the field configuration in case of inverse power-law potentials
in Minkowski space time.  

Heuristically speaking, expansion generally works against caustic formation
and its effects becomes crucial for inverse power-law potentials with
$~0<n<2$. If
the potential vanishes faster than $1/\phi^2$ at infinity,  the dust like solution is late time attractor;
the exponentially decreasing potential belongs to this category. In this case, caustics form
in Minkovski space very distinctly. The cosmic expansion does dilute the effect of caustic formation
but can not irradicate  them. 
 It is really interesting that  for  the inverse power law potentials under consideration, the effect of expansion can compete with the tendency of caustic formation.  Indeed,  dark energy as a late time attractor
 of the dynamics in this case,  gives rise to cosmic repulsion allowing to avoid caustics and
multi-valued regions in the field profile.

From the behavior of the scalar field with exponentially increasing
and inverse power-law potentials, we infer that field first rolls
fast and mimics dark matter and subsequently gives rise to dark
energy, see figure \ref{fig13}. Our simulation shows that
the particle trajectories computed in the expanding universe using
the exact evolution equations (18) and (19) broadly agree with result
obtained assuming dark energy dominance at all times.

Evolution of field configurations at different times is similar for
the cases of exponentially decreasing rolling massive and inverse
power-law potentials apart from the initial and transition periods
in the Minkowski space and the expanding universe. It should be
noted that in the case of the expanding universe, the pattern of the
evolution of the field remains almost same as its initial profile
for the scalar field dominated universe. Since the field is almost
steady for the exponentially increasing rolling massive scalar
potential, the field configuration is same as its initial profile.
Thus taking into account the results of Ref.~\cite{staro}, we
conclude that caustics generally form for DBI
systems in Minkowski space time with an exception of massive rolling scalar. 
On the other
hand in  FRW expanding universe, caustics form in DBI systems
with potentials decaying faster than $1/\phi^2$ at infinity, in particular
 the exponentially decreasing rolling massive scalar potential. As for  the
inverse power-law potentials under consideration, they are free from
caustics and may be suitable to explain the late time cosmic
acceleration.

\section*{Acknowledgments}
We are indebted to A. Starobinsky for his valuable comments. We also thank S. 
Tsujikawa for useful discussion. UDG  thanks the Center for Theoretical 
Physics, Jamia Millia Islamia, New Delhi for  hospitality during his visit. MS 
is supported by DST project No.SR/S2/HEP-002/2008. MS thanks ICTP for 
hospitality.

\end{document}